\documentclass[twocolumn]{autart}    
                                   
\usepackage[colon]{natbib}
\usepackage {indentfirst}
\usepackage{algorithm}
\usepackage{algorithmicx} 
\usepackage{algpseudocode}

\usepackage{amsmath}

\usepackage{amsfonts}
\usepackage{mathrsfs}
\usepackage{mathtools}
\usepackage{mathdots}
\usepackage{textcomp}
\usepackage{tikz}
\usepackage{booktabs}
\usepackage{graphicx}          
\usepackage{bm} 

\usepackage{cases} 
\usepackage {amssymb}
\usepackage{appendix}
\usepackage{caption}
\usepackage{tablefootnote} 
\usepackage{hyperref} 

\renewcommand{\vec}{\mbox{vec}}
\newcommand{\rank}{\mbox{rank}}

\newcommand*{\circled}[1]{\lower.7ex\hbox{\tikz\draw (0pt, 0pt)%
		circle (.42em) node {\makebox[1em][c]{\small #1}};}}
\allowdisplaybreaks[4]

\newtheorem{definition}{Definition}

\newtheorem{assumption}{Assumption}

\newtheorem{theorem}{Theorem}

\newtheorem{remark}{Remark}

\newtheorem{proof}{Proof}  
 
 \newtheorem{example}{Example}       
  
 \newtheorem{lemma}{Lemma}         
    
  \newtheorem{corollary}{Corollary}      
\begin{document}

\begin{frontmatter}

\title{Informativity Conditions for Multiple Signals: Properties, Experimental Design, and Applications (extended version)\thanksref{footnoteinfo}} 

\thanks[footnoteinfo]{This work was supported by the National Natural Science Foundation of China under Grant 62103203. (Corresponding author: Fuyong Wang.)}
\author[Tianjin]{Ao Cao}\ead{caoao@mail.nankai.edu.cn},  
\author[Tianjin]{Fuyong Wang}\ead{wangfy@nankai.edu.cn}

\address[Tianjin]{College of Artificial Intelligence, Nankai University, Tianjin 300350, China} 


\begin{abstract}                          
	Recent studies highlight the importance of the persistently exciting condition in a single signal sequence for model identification and data-driven control techniques. However, maintaining prolonged excitation in control signals introduces significant challenges, as continuous excitation can reduce the lifetime of mechanical devices. In this paper, we introduce three informativity conditions for various types of multi-signal data, each augmented with weight factors. We explore the interrelations between these conditions and their rank properties in linear time-invariant systems. All three conditions can extend Willems’ fundamental lemma and are utilized to assess the properties of the system. Furthermore, we introduce open-loop experimental design methods tailored to each of the three conditions, which can synthesize the required excitation conditions either offline or online, even in the presence of limited information within each signal segment.  Illustrative examples confirm that these conditions yield satisfactory outcomes in both least-squares identification and the construction of data-driven controllers. 
\end{abstract}

\end{frontmatter}

\section{Introduction}
With the continuous development and application of data-driven technologies, the academic interest in data informativity has surged notably. In this context, the concept of persistently exciting (PE) condition \citep{WILLEMS2005325} plays a crucial role. The PE condition requires that a signal be sufficiently rich over time to capture all dynamic characteristics of the system. This requirement is fundamental to various system identification methods \citep{ioannou2006adaptive,katayama2005subspace,MARKOVSKY2005755,markovsky2012low,LOVERA20001639} and data-driven control methods \citep{9109670,9744574,9903320,8933093,PARK20091265,10124991}.

While persistency of excitation can indeed be enforced in theory by well-known inputs  \citep{5325719} such as pseudo-random binary sequences (PRBS), broadband white noise, or periodic signals with many harmonics, their practical deployment is often constrained. First, guaranteeing PE with these signals typically requires a long single trajectory, which may not be available in experimental environments where data collection must be interrupted or restarted. Second, long-duration stochastic or high-frequency excitations can violate safety or performance constraints in many systems, e.g., process reactors or power electronic converters where sustained probing may be unsafe \citep{YANG2024100986}, or robotic platforms where prolonged excitation induces mechanical wear \citep{doi:10.1177/0278364919853618}. Third, when multiple shorter trajectories are collected, naive juxtaposition of PRBS-type signals does not necessarily yield a full-rank Hankel matrix due to phase misalignment or heterogeneous signal magnitudes. These limitations highlight the need for conditions that certify data informativity beyond the classical single-trajectory PE framework.

Existing research has shown that the PE condition is not a necessary prerequisite for system identification and data-driven control design \citep{8960476, KANG2023111130,10317631}. In the domain of system identification, methods such as dynamic regressor extension and mixing \citep{7526771, 9121700} have relaxed the PE condition by extending the regression model in the time domain. Certain adaptive methods \citep{DHAR2022100672} only require that excitation conditions be met during the initial phase. Furthermore, the challenge of insufficient information from a single sensor has been addressed through distributed frameworks \citep{W6578135, doi:10.1137/16M1106791} and centralized frameworks \citep{7098430}. Despite these advancements, these methods often relax the information content of system regression vectors or system trajectories, yet rarely focus on designing from the perspective of control signals. Therefore, a clear experimental design strategy for system input signals remains essential. While current experimental design theories are primarily based on the premise of satisfying identifiability \citep{DEPERSIS2021285,9406124,BOMBOIS20061651,HJALMARSSON2005393}, there remains a significant gap in research regarding how to design experiments that guarantee the effectiveness of identification methods when the informativeness of input signals is lacking. In such cases, the use of multiple trajectories may be necessary.

In the design of data-driven control methods, a new control framework is developed in \citep{8960476} that eliminates the stringent requirement of PE, but the trajectory data used still needs to guarantee the identifiability condition. The collectively persistently exciting (CPE) condition proposed in \citep{9062331} and \citep{9682952} offers potential insights for the design of data-driven control methods using trajectories generated by insufficiently informative controllers. By organizing the Hankel matrices that evaluate the excitability of each signal sequence side-by-side into a larger rank judgment matrix, this condition extends the fundamental lemma from a single signal sequence to multiple signal sequences. However, this large matrix structure can lead to a dramatic increase in computational burden as the number of columns increases, and excessive differences in the magnitude of the values between each Hankel matrix can also cause deviations from the expected results. Additionally, it is currently an open question how to provide more choices of matrix structures for different data-driven problems \citep{MARKOVSKY202142}.

Motivated by the aforementioned challenges, the objective of this paper is to relax the PE condition typically required for a single control sequence in data-driven methods, while offering more flexible matrix structure choices for a variety of data-driven problems. Furthermore, we aim to design experimental methodologies that ensure the effectiveness of data-driven techniques even in scenarios where data informativeness is lacking.  Specifically, the contributions of this paper are summarized as follows:
\begin{itemize}
\item	
We introduce three weighted CPE conditions for multiple signal sequences with different dimensions, same dimensions, and partially the same dimensions. All three conditions relax the PE condition requirement for a single signal sequence. Furthermore, we analyze the feasibility of implementing these conditions and their transformation relationships, and provide an initial exploration of which CPE conditions are most suitable for different problem types.
\item	
For the three CPE conditions, we design experimental schemes and demonstrate that these experimental designs remain effective even when signal sequence information is insufficient. We also explore how control sequences satisfying the CPE conditions guide the convergence precision of least-squares (LS) estimators. Furthermore, by utilizing the three CPE conditions, we extend Willems' fundamental lemma, and the extended lemma yields satisfactory results in data-driven control methods.
\end{itemize}

The remainder of this paper is organized as follows. Section \ref{s2} introduces the preliminaries, covering key definitions and a review of prior work on the PE condition.  In Section \ref{s3}, we define three CPE conditions tied to data dimensions, analyze their transformational relationships and rank properties, and extend Willems' fundamental lemma.  Section \ref{s4} introduces experimental methods tailored to the three CPE conditions and demonstrates their validity.  Section \ref{s6} provides illustrative examples of applications to validate the methodologies and theories discussed, while Section \ref{s7} concludes the paper.
\section{Preliminaries}\label{s2}
\subsection{Notation}
Throughout this paper, $\mathbb{Z}$, $\mathbb{N}$, $\mathbb{R}$ and $\mathbb{R}^{m\times n}$ denote the sets of integers, natural numbers, real numbers and $m \times n$ real matrices, respectively. $0_{m\times n}$ denotes a zero matrix with $m$ rows and $n$ columns. $\bm{1}_n$ represents the $n$-dimensional column vector with all entries set to one. $I_n$ and $0_n$ refer to the identity matrix and zero matrix of order $n$, respectively, without subscripts to indicate the suitable dimension. The modulo operation $\text{mod}(a, b)$ returns the remainder when $a$ is divided by $b$.

For a real matrix $P \in \mathbb{R}^{m \times n}$, let $P_{vec} = [P_1, P_2, \ldots, P_m]^T$, where $P_i$ represents the $i$-th row of $P$. The right kernel and left kernel of $P$ refer to the spaces of all real column vectors $v_1$ and row vectors $v_2$, respectively, satisfying $Pv_1 = 0$ and $v_2P = 0$. The image, right kernel, left kernel, rank and Moore–Penrose pseudoinverse of $P$ are denoted as $\text{im}\,P$, $\text{ker}\,P$, $\text{leftker}\,P$, $\text{rank}\,P$ and $P^{\dag}$, respectively. Additionally, let $\lambda_{max}(P)$, $||P||$, and $||P||_2$ represent the maximal eigenvalue, Frobenius norm, and spectral norm of $P$, respectively. 

Given a signal $z$: $\mathbb{Z}\rightarrow \mathbb{R}^{m}$, we define $z^{[k,k+T_0]}$ as $$z^{[k,k+T_0]}=\begin{bmatrix}
{z(k)}\\
\vdots \\
{z(k + {T_0})}
\end{bmatrix},$$
where $k\in \mathbb{Z}$ and $T_0\in\mathbb{N}$. The block Hankel
matrix of order $L$ associated to $z^{[k,k+T_0]}$ is defined as
\begin{align*}
&{H_L}({z^{[k,k + {T_0}]}})\\
&~~=\begin{bmatrix}
	{z(k)}&{z(k + 1)}& \cdots &{z(k + {T_0} - L + 1)}\\
	{z(k + 1)}&{z(k + 2)}& \cdots &{z(k + {T_0} - L + 2)}\\
	\vdots & \vdots & \ddots & \vdots \\
	{z(k + L - 1)}&{z(k + L)}& \cdots &{z(k + {T_0})}
\end{bmatrix}.&
\end{align*}
\subsection{Persistency of Excitation for a Single Sequence}
The concept of persistency of excitation for a single signal sequence, as defined in \cite{WILLEMS2005325}, is elucidated below.

\begin{definition}( \citep{WILLEMS2005325}).\label{Persistently Exciting Condition}
A signal sequence $z^{[0,T_0-1]}$ with $z(k)\in \mathbb{R}^{m}$, $k=0, ..., T_0-1$ is PE of order $L$ if the block Hankel matrix $H_L(z^{[0,T_0-1]})$ has full row rank $mL$. 
\end{definition}

It is intuitive from this definition that a PE signal of order $L$ possesses the ability to span the entire real space $\mathbb{R}^{mL}$. In particular, if the input signal to a controllable linear time-invariant (LTI) system exhibits sufficiently high-order persistency of excitation, some attractive properties arise. Consider the LTI system 
\begin{align}\label{Sysform. (1)}
x(k+1) = Ax(k) + Bu(k),
\end{align}
where $A \in {\mathbb{R}^{n \times n}}$, ${B} \in {\mathbb{R}^{n \times {m}}}$, ${x(k)} \in {\mathbb{R}^n}$ and ${u(k)} \in {\mathbb{R}^{m}}$ represent the state vector and control input vector, respectively. The pair $(A,B)$ is controllable. A fundamental property in \citep[Cor. 2]{WILLEMS2005325} demonstrates that the LTI system \eqref{Sysform. (1)} satisfies the rank condition
\begin{align}\label{Rank condition. (1)}
\text{rank}\,\begin{bmatrix}
	{{H_1}({x^{[0,{T_0} - L]}})}\\
	{{H_L}({u^{[0,{T_0} - 1]}})}
\end{bmatrix} = n + mL
\end{align}
if the control input $u^{[0,{T_0} - 1]}$ is PE of order $L+n$.  

In addition, the rank condition \eqref{Rank condition. (1)} subtly implies that the sequences ${{{\bar x}^{[0,L - 1]}}}$ and ${{{\bar u}^{[0,L - 1]}}}$ form an $L$-long state-input trajectory of \eqref{Sysform. (1)} if and only if there exists $g\in\mathbb{R}^{T_0-L+1}$ such that
\begin{align}\label{Fundamental lemma. (1)}
\begin{bmatrix}
	{{{\bar x}^{[0,L - 1]}}}\\
	{{{\bar u}^{[0,L - 1]}}}
\end{bmatrix} = \begin{bmatrix}
	{{H_L}({x^{[0,{T_0} - 1]}})}\\
	{{H_L}({u^{[0,{T_0} - 1]}})}
\end{bmatrix}g. 
\end{align}

This property indicates that the image of the matrix $\begin{bmatrix}
{{H_L}({x^{[0,{T_0} - 1]}})}\\
{{H_L}({u^{[0,{T_0} - 1]}})}
\end{bmatrix}$ is the representation of the $L$-long state-input trajectory space of the LTI system \eqref{Sysform. (1)}. 
It has been rigorously proved in the behavioral system framework \citep[Th. 1]{WILLEMS2005325} and the state-space framework \citep[Lem. 2]{8933093}, respectively. 

The properties described by \eqref{Rank condition. (1)} and \eqref{Fundamental lemma. (1)}, often referred to as Willems’ fundamental lemma due to their foundational importance for system identification and data-driven control. 

Indeed, if the input sequence $u^{[0,{T_0} - 1]}$ of system \eqref{Sysform. (1)} is PE of order $n + 1$, the matrix $\begin{bmatrix}
{{H_1}({x^{[0,{T_0} - 1]}})}\\
{{H_1}({u^{[0,{T_0} - 1]}})}
\end{bmatrix}$ has full row rank.  This condition enables the recovery of the system matrices $A$ and $B$ from the data $x^{[0,{T_0} - 1]}$ and $u^{[0,{T_0} - 1]}$, i.e.,  identifiability condition (see \citep{9904308,8960476}).  The system matrices can be uniquely determined by solving the following equation:
\begin{align}\label{single-system identification.(2)}
\begin{bmatrix}
	A&B
\end{bmatrix}={H_1}({x^{[1,{T_0}]}}){\begin{bmatrix}
		{{H_1}({x^{[0,{T_0} - 1]}})}\\
		{{H_1}({u^{[0,{T_0} - 1]}})}
	\end{bmatrix}^\dag}.
	\end{align}
	
	On the other hand, under the influence of the prior control input $ u^{[0,T_0-1]}$ that satisfies the PE condition, the system trajectories ${\bar x}^{[0,L-1]}$ and ${\bar u}^{[0,L-1]}$ can form the non-parametric representation in \eqref{Fundamental lemma. (1)}, which solely relies on the prior data. Consequently, in addition to its important role in system identification, the PE condition has received widespread attention in the development of data-driven controllers such as parametric model reconstruction methods \citep{8933093,10124991} and data-driven model predictive control methods \citep{9109670,9744574,9903320}.
	
	Even though it has been shown how to ensure the informativeness of a single control sequence \citep{5325719}, in some cases it may still not be possible to generate sufficiently informative trajectories due to the many limitations of the experiment.  However, designing experiments with insufficiently informative controllers fails to ensure the identifiability condition and the feasibility of data-driven approaches.
	
	\begin{example}
When applying a  feedback controller $u = Kx$, where $K$ is the control gain, we have 
$$ {{H_L}({u^{[0,{T_0} - 1]}})} = \begin{bmatrix}
	{K{{(A + BK)}^0}{H_1}({x^{[0,{T_0} - L]}})}\\
	\vdots \\
	{K{{(A + BK)}^{L - 1}}{H_1}({x^{[0,{T_0} - L]}})}
\end{bmatrix},$$
which means that every row of ${{H_L}({u^{[0,{T_0} - 1]}})}$ is in the row space of ${H_1}({x^{[0,{T_0} - L]}})$. Therefore,
$$\text{rank}\ {{H_L}({u^{[0,{T_0} - 1]}})}<mL.$$
Consequently,
$$  \text{rank}\,\begin{bmatrix}
	{{H_1}({x^{[0,{T_0} - L]}})}\\
	{{H_L}({u^{[0,{T_0} - 1]}})}
\end{bmatrix} \leq n$$
and rank condition \eqref{Rank condition. (1)} is not satisfied. In this case, neither  identifiability nor fundamental lemma can be guaranteed.
\end{example}

This raises three natural questions. First, can multiple non-persistently exciting sequences be employed to relax the requirement for persistent excitation of a single sequence as defined in Definition \ref{Persistently Exciting Condition}, and if so, how can they be utilized? Second, how can we design experiments to address the first question? Third, is it feasible to leverage multiple non-persistently exciting control sequences for model identification and data-driven controller design?  These three questions are the subject of the remainder of this paper. 

\section{Three Collectively Persistently Exciting Conditions and Their Properties}\label{s3}

In this section, we introduce three collectively persistently exciting (CPE) conditions for multi-trajectory data, aiming to eliminate the requirement of persistently excited single-trajectory sequence as specified in \eqref{Rank condition. (1)}. These conditions are delineated from three perspectives: same-dimension data, different-dimension data, and partially same-dimension data. Theorem  \ref{lem.transformative} elucidates the transformation relations among these CPE conditions, while Lemma \ref{lem.CPE.Rank} establishes a rank condition akin to \eqref{Rank condition. (1)}, and Lemma \ref{lem.extend} extends Willems' fundamental lemma under CPE conditions. 

\begin{figure}[htbp]
\centering
\includegraphics[scale=0.65]{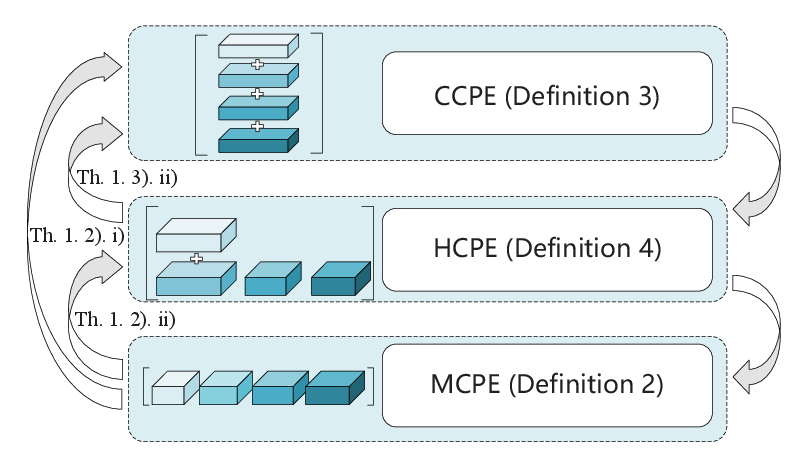}
\caption {{Transformative relationship between the three CPE conditions. The left side of the figure illustrates the synthesis of the conditions, while the right side presents the specific conditions. The arrows denote the derived relationships.}}
\label{CPE}
\end{figure}
\subsection{Three Collectively Persistently Exciting Conditions}
In this subsection, we delineate the CPE conditions into three types: cumulative, mosaic, and hybrid, tailored for multi-trajectory sequences. Unlike the traditional PE condition, where each trajectory needs to exhibit sufficient excitation and satisfy identifiable conditions, our approach shifts focus to the collective impact. The three distinct matrix structures afford greater flexibility for trajectories of varying lengths, thereby catering to a wide array of problem types.

Before proceeding, let $\{z_{i}^{[k,k+T_i]}\}_{i=1}^p$ represent the sequences $z_{1}^{[k,k+T_1]}, z_{2}^{[k,k+T_2]}, \ldots, z_{p}^{[k,k+T_p]}$, where  $z_{i}^{[k,k+T_i ]}$ denotes the value of the $i$-th signal $z_i$ within a specific time interval $[k,k+T_i ]$, with $T_i$  representing a natural number associated with index $i$. 

We begin by introducing a mosaic collectively persistently exciting (MCPE) condition  for signal sequences of different dimensions, which is shown to be the easiest exciting condition to satisfy in the subsequent discussion (see Theorem \ref{lem.transformative}).
\begin{definition}(Collectively Persistently Exciting Condition (Type-\uppercase\expandafter{\romannumeral1}): mosaic).\label{CPE-2}
Multiple signals $\{z^{[0,T_i-1]}_i\}_{i=1}^{p}$ with $z_{i}(k)\in \mathbb{R}^{m}$, $k= 0, 1, ... ,{T_i-1}$ are mosaic collectively persistently exciting of order $L$ if the matrix 
\begin{align*}
	&{H_L^{mos}(\{z^{[0,T_i-1]}_i\}_{i=1}^{p})}\\ \nonumber
	&~~~~~~~~~~~~~~~~~~~={\begin{bmatrix}
			{{\alpha _1}{H_L}({z_{1}^{[0,{T_1} - 1]}})}\  \ldots \ {{\alpha _p}{H_L}({z_{p}^{[0,{T_p} - 1]}})}
	\end{bmatrix}}
\end{align*}
has full row rank $mL$ for any $\alpha _i \neq 0$. 
\end{definition}

Compared to the PE condition for a single trajectory, the MCPE condition is motivated by practical scenarios where generating one sufficiently long informative trajectory is infeasible. Typical examples include process systems constrained by safety limits (e.g., reactors or power electronics that cannot be excited for long durations), robotic platforms where prolonged excitation causes mechanical wear, or networked systems with limited storage or communication. In such cases, multiple short (non-PE) trajectories are far easier to obtain.

The MCPE condition does not specify the number of dimensions of each signal sequence, i.e., the $T_i$ of each signal can be different. Only the sum of all signal lengths is required to satisfy $\sum\limits_{i = 1}^p {{T_i}} \ge mL + p(L - 1)$. However, this flexibility results in an unrestricted number of columns in the rank test matrix ${H_L^{mos}(\{z^{[0,T_i-1]}_i\}_{i=1}^{p})}$, potentially leading to an excessive number of redundant columns in the matrix. This abundance of columns can significantly increase computational complexity in practical applications such as model predictive control \cite{9109670,9744574,9903320}.

Therefore, in the following definition, we introduce a cumulative collectively persistently exciting (CCPE) condition for same-dimensional signal sequences. This condition is subsequently confirmed to be the most challenging to satisfy but is the most suitable in terms of computational complexity for application in the design of data-driven control methods.
\begin{definition}(Collectively Persistently Exciting Condition (Type-\uppercase\expandafter{\romannumeral2}): cumulative).\label{CPE-1}
Multiple signals $\{z^{[0,T_0-1]}_i\}_{i=1}^{p}$ with $z_{i}(k)\in \mathbb{R}^{m}$, $k= 0, 1, ... ,{T_0-1}$ are cumulative collectively persistently exciting  of order $L$ if the matrix 
\begin{align*}
	{H_L^{cum}(\{z^{[0,T_0-1]}_i\}_{i=1}^{p})}=\sum\limits_{i = 1}^p {{\alpha _i}{H_L}({z^{[0,T_0 - 1]}_{i}})} 
\end{align*}
has full row rank $mL$ for any $\alpha _i \neq 0$. 
\end{definition}

The CCPE condition strictly demands that all signal sequences share the same dimension/length, with a minimum requirement of ${T_0} \ge (m+1)L  - 1$. This stringent dimensionality requirement for different signal sequences ensures that the rank-judgment matrix ${H_L^{cum}(\{z^{[0,T_0-1]}_i\}_{i=1}^{p})}$ can be confined to a relatively smaller size, specifically, $mL\times {(T_0-L+1)}$ dimensions. Consequently, when it comes to computations using rank judgment matrices, the CCPE condition can better restrict the complexity of the matrices compared to the MCPE condition. This restraint effectively serves the purpose of reducing computational costs.

Next, we introduce a hybrid collectively persistently exciting (HCPE) condition for partially same-dimensional signal sequences. This condition employs a rank test matrix that combines the matrix structures from both the CCPE condition and the MCPE condition, leveraging the advantages and mitigating the disadvantages of each.
\begin{definition}(Collectively Persistently Exciting Condition (Type-\uppercase\expandafter{\romannumeral3}): hybrid).\label{CPE-3}
Multiple signals $\{z^{[0,T_0-1]}_i\}_{i=1}^{\bar{p}}$ and $\{z^{[0,T_i-1]}_i\}_{i=\bar{p}+1}^{p}$ with $z_{i}(k)\in \mathbb{R}^{m}$, $k= 0, 1, ... ,{T_i-1}$  are hybrid collectively persistently exciting of order $L$ if the matrix 
\begin{align*}
	&{H_L^{hyb}(\{z^{[0,T_i-1]}_i\}_{i=1}^{p})}\\ \nonumber
	&~~~~~~={\begin{bmatrix}
			{H_L^{cum}(\{z^{[0,T_0-1]}_i\}_{i=1}^{\bar{p}})} &{H_L^{mos}(\{z^{[0,T_i-1]}_i\}_{i=\bar{p}+1}^{p})}
	\end{bmatrix}} 
\end{align*}
has full row rank $mL$ for any $\alpha _i \neq 0$. 
\end{definition}

The HCPE condition requires the first $\bar{p}$ signal sequences (which can be any $\bar{p}$ sequences, without loss of generality, and are taken here as the first $\bar{p}$) to be of the same dimension/length, with $T_0 + \sum\limits_{i = \bar{p}}^p {{T_i}} \ge mL + (p-\bar{p}+1)(L - 1)$.

It is important to note that none of the three CPE conditions require each signal sequence to be PE. Rather, they allow for insufficient information sequences but mandate that the overall behavior demonstrates a PE-like performance, which is physically achievable. In Section \ref{s4}, we propose experimental methods to synthesize the three CPE conditions in cases where each signal segment is not PE.

\begin{remark}
The benefit of MCPE lies not in reducing the number of matrix columns, but in simplifying the overall computational procedure of informativity verification while enhancing flexibility in handling heterogeneous trajectories. Instead of repeatedly checking PE on each trajectory or artificially aligning all trajectories to the same horizon (as required by CCPE), MCPE organizes all heterogeneous trajectories into a single rank test. Thus, only one unified check is needed, while weighting factors mitigate numerical ill-conditioning caused by magnitude discrepancies across datasets. Although the resulting matrix may be wider, the computation pipeline is considerably simpler  in certain scenarios, especially when handling heterogeneous-length data, which is shown with the illustrative example in the Section \ref{complexity discussion}. The CPE condition proposed in \cite{9062331} shares similarities with the first type introduced in this paper, the MCPE condition, as both combine Hankel matrices in a side-by-side fashion. However, the MCPE condition distinguishes itself by incorporating weight factors, $\alpha_i$, which allows it to perform better when handling data with significant order-of-magnitude differences.  Additionally, a careful selection of the weighting factors can enhance both performance and computational efficiency when employing data-driven methods, as demonstrated in the examples in Section \ref{s6.A}. Notably, despite the introduction of weight factors, the difficulty of synthesis remains unchanged. This is attributed to the fact that for any $\alpha_i \neq 0$, the rank of the matrix ${H_L^{mos}(\{z^{[0,T_i-1]}_i\}_{i=1}^{p})}$ remains the same as when $\alpha _i =1$.
\end{remark}

\subsection{Transformation Relations and rank properties of CPE Conditions}
In this subsection, we analyze the transformation relations between the three CPE conditions. The result reflects the difficulty of realizing the three conditions, and provides some theoretical basis for the selection of the three conditions in practical identification and control problems.
\begin{theorem}\label{lem.transformative}
For multiple signals $\{z^{[0,T_i-1]}_i\}_{i=1}^{p}$ with $z_{i}(k)\in \mathbb{R}^{m}$, $k= 0, 1, ... ,{T_i-1}$, the following transformations hold:
\begin{itemize}
	\item[$1).$] 
	If $\{z_{i}^{[0,{T_i-1}]}\}_{i=1}^p$ are CCPE $\Longrightarrow$ $\{z_{i}^{[0,{T_i-1}]}\}_{i=1}^p$ are also MCPE and HCPE; 
\end{itemize}
\begin{itemize}
	\item[$2).$] 
	If $\{z_{i}^{[0,{T_i-1}]}\}_{i=1}^p$ are MCPE and satisfy
	\begin{itemize}
		\item[\romannumeral1).] 
		$T_0=T_1=...=T_{p}$, $\text{im} \, {{H_L^{mos}(\{z^{[0,T_i-1]}_i\}_{i=1}^{p})}^\top}
		\cap \text{leftker}\, {\textbf{1}_p} \otimes {I_{{T_0} - L + 1}}=\{0\}$ $\Longrightarrow$ $\{z_{i}^{[0,{T_i-1}]}\}_{i=1}^p$ are also CCPE;
		\item[\romannumeral2).] 
		$T_0=T_1=...=T_{\bar{p}}$, $\text{im}\,  {{H_L^{mos}(\{z^{[0,T_i-1]}_i\}_{i=1}^{\bar{p}})}^\top}
		\cap \text{leftker}\, {\begin{bmatrix}
				{{\textbf{1}_{\bar p}}}&{{0_{\bar p \times (p - \bar p)}}}\\
				{{0_{p - \bar p}}}&{{I_{p - \bar p}}}
		\end{bmatrix}} \otimes {I_{{T_0} - L + 1}}=\{0\}$ $\Longrightarrow$ $\{z_{i}^{[0,{T_i-1}]}\}_{i=1}^p$ are also HCPE;
	\end{itemize}	
\end{itemize}
\begin{itemize}
	\item[$3).$] 
	If $\{z_{i}^{[0,{T_i-1}]}\}_{i=1}^p$ are HCPE, 
	\begin{itemize}
		\item[\romannumeral1).] 
		$\{z_{i}^{[0,{T_i-1}]}\}_{i=1}^p$ are also MCPE;
		\item[\romannumeral2).] 
		$T_0=T_{\bar{p}+1}=...=T_{p}$, $\text{im}\,  {{H_L^{hyb}(\{z^{[0,T_i-1]}_i\}_{i=1}^{p})}^\top}
		\cap \text{leftker}\, {\textbf{1}_{\bar{p}+1}} \otimes {I_{{T_0} - L + 1}}=\{0\}$ $\Longrightarrow$ $\{z_{i}^{[0,{T_i-1}]}\}_{i=1}^p$ are also CCPE.
	\end{itemize}	
\end{itemize}
\end{theorem}
\textbf {Proof.}  See the Appendix. A. \hfill $\blacksquare$ 

Theorem \ref{lem.transformative} demonstrates  that the MCPE condition is the most straightforward to achieve, followed by the HCPE condition, with the CCPE condition being the most difficult. A significant finding is that satisfying both the CCPE and HCPE conditions results in fulfilling the MCPE condition. Fig. \ref{CPE} illustrates the transformation relationship among the three conditions.

In the following lemma, we establish the rank condition under three CPE conditions, corresponding to the rank condition \eqref{Rank condition. (1)}. This lemma is pivotal in extending Willems' fundamental lemma (see Lemma \ref{lem.extend}).

\begin{lemma}\label{lem.CPE.Rank}
Assuming $\{u_{i}^{[0,{T_i-1}]},x_{i}^{[0,{T_i-1}]}\}_{i=1}^p$ constitutes the $p$-segment input-state trajectory of the system \eqref{Sysform. (1)}.
\begin{itemize}
	\item[$1).$] 
	If $T_0=T_1=T_2=...=T_{p}$ and $\{u_{i}^{[0,{T_0-1}]}\}_{i=1}^p$ are CCPE of order $L+n$, then $$\text{rank}\begin{bmatrix}
		{H_1^{cum}(\{{x_{i}^{[0,{T_i-L}]}\}_{i=1}^p})}\\
		{H_L^{cum}(\{{u_{i}^{[0,{T_i-1}]}\}_{i=1}^p})}
	\end{bmatrix} = n + mL.$$
	
	\item[$2).$] 
	If $\{u_{i}^{[0,{T_i-1}]}\}_{i=1}^p$ are MCPE of order $L+n$, then $$\text{rank}\begin{bmatrix}
		{H_1^{mos}(\{{x_{i}^{[0,{T_i-L}]}\}_{i=1}^p})}\\
		{H_L^{mos}(\{{u_{i}^{[0,{T_i-1}]}\}_{i=1}^p})}
	\end{bmatrix}= n + mL.$$

	\item[$3).$] 
	If $T_0=T_1=T_2=...=T_{\bar{p}}$ and $\{u_{i}^{[0,{T_i-1}]}\}_{i=1}^p$ are HCPE of order $L+n$, then $$\text{rank}\begin{bmatrix}
		{H_1^{hyb}(\{{x_{i}^{[0,{T_i-L}]}\}_{i=1}^p})}\\
		{H_L^{hyb}(\{{u_{i}^{[0,{T_i-1}]}\}_{i=1}^p})}
	\end{bmatrix}= n + mL.$$
\end{itemize}
\end{lemma}
\textbf {Proof.}  See the Appendix. B. \hfill $\blacksquare$ 

\subsection{Extension of Fundamental Lemma }
In this subsection, we delve into the fundamental lemmas under the CPE condition, extending the traditional Willems' lemma. 

For a single-input signal sequence satisfying the PE condition of order $L + n$, \eqref{Fundamental lemma. (1)} always holds. A similar conclusion applies when the signal sequence extends to more than one.
\begin{lemma}\label{lem.extend}
Assuming $\{u_{i}^{[0,{T_i-1}]},x_{i}^{[0,{T_i-1}]}\}_{i=1}^p$ constitutes the $p$-segment input-state trajectory of the system \eqref{Sysform. (1)}. Then, the following holds. 
\begin{itemize}
	\item[$1).$] 
	If $T_0=T_1=T_2=...=T_{p}$ and $\{u_{i}^{[0,{T_i-1}]}\}_{i=1}^p$ are CCPE of order $L+n$, then any $L$-long input-state trajectory of system \eqref{Sysform. (1)} can be expressed as  $$\begin{bmatrix}
		{H_L^{cum}(\{{x_{i}^{[0,{T_i-1}]}\}_{i=1}^p})}\\
		{H_L^{cum}(\{{u_{i}^{[0,{T_i-1}]}\}_{i=1}^p})}
	\end{bmatrix}g,$$
	where $g$ is a real vector.
	\item[$2).$] 
	If $\{u_{i}^{[0,{T_i-1}]}\}_{i=1}^p$ are MCPE of order $L+n$, then any $L$-long input-state trajectory of system \eqref{Sysform. (1)} can be expressed as
	$$\begin{bmatrix}
		{H_L^{mos}(\{{x_{i}^{[0,{T_i-1}]}\}_{i=1}^p})}\\
		{H_L^{mos}(\{{u_{i}^{[0,{T_i-1}]}\}_{i=1}^p})}
	\end{bmatrix}g,$$
	where $g$ is a real vector.
	\item[$3).$] 
	If $T_0=T_1=T_2=...=T_{\bar{p}}$ and $\{u_{i}^{[0,{T_i-1}]}\}_{i=1}^p$ are HCPE of order $L+n$, then any $L$-long input-state trajectory of system \eqref{Sysform. (1)} can be expressed as
	$$\begin{bmatrix}
		{H_L^{hyb}(\{{x_{i}^{[0,{T_i-1}]}\}_{i=1}^p})}\\
		{H_L^{hyb}(\{{u_{i}^{[0,{T_i-1}]}\}_{i=1}^p})}
	\end{bmatrix}g,$$
	where $g$ is a real vector.
\end{itemize}
\end{lemma}
\textbf {Proof.}  By the dynamics of system \eqref{Sysform. (1)}, one has
\begin{align}\label{Behavior.proof1}
&\begin{bmatrix}
	{H_L^{cum}(\{{u_{i}^{[0,{T_0-1}]}\}_{i=1}^p})}\\
	{H_L^{cum}(\{{x_{i}^{[0,{T_0-1}]}\}_{i=1}^p})}
\end{bmatrix}g \nonumber \\ 
&~~~~~~~~~~~~~~= \begin{bmatrix}
	I&0\\
	\mathcal{T}_L&\mathcal{O}_L
\end{bmatrix}\begin{bmatrix}
	{H_L^{cum}(\{{u_{i}^{[0,{T_0-1}]}\}_{i=1}^p})}\\
	{H_1^{cum}(\{{x_{i}^{[0,{T_0-L}]}\}_{i=1}^p})}
\end{bmatrix}g,
\end{align}
where $\mathcal{T}_L$ and $\mathcal{O}_L$ are defined as
$${{\mathcal T}_L} = \begin{bmatrix}
0&0&0& \cdots &0\\
B&0&0& \cdots &0\\
{AB}&B&0& \cdots &0\\
\vdots & \vdots & \vdots & \ddots & \vdots \\
{{A^{L - 2}}B}&{{A^{L - 2}}B}&{{A^{L - 2}}B}& \cdots &0
\end{bmatrix}$$
$${{\mathcal O}_L} = {\begin{bmatrix}
	I&{{A^\top}}&{{{({A^2})}^\top}}& \cdots &{{{({A^{L - 1}})}^\top}}
\end{bmatrix}^\top}.$$

Since $\begin{bmatrix}
{H_L^{cum}(\{{u_{i}^{[0,{T_0-1}]}\}_{i=1}^p})}\\
{H_1^{cum}(\{{x_{i}^{[0,{T_0-L}]}\}_{i=1}^p})}
\end{bmatrix}$ has full row rank by Lemma \ref{lem.CPE.Rank}, all vectors consisting of the initial state $\bar{x}(0)$ and the $L$-long input sequence $\bar{u}^{[0,L-1]}$ of system \eqref{Sysform. (1)} can be represented by a linear combination of the columns of $\begin{bmatrix}
{H_L^{cum}(\{{u_{i}^{[0,{T_0-1}]}\}_{i=1}^p})}\\
{H_1^{cum}(\{{x_{i}^{[0,{T_0-L}]}\}_{i=1}^p})}
\end{bmatrix}$, i.e., $$\begin{bmatrix}
{H_L^{cum}(\{{u_{i}^{[0,{T_0-1}]}\}_{i=1}^p})}\\
{H_1^{cum}(\{{x_{i}^{[0,{T_0-L}]}\}_{i=1}^p})}
\end{bmatrix}g=\begin{bmatrix}
{{{\bar u}^{[0,L - 1]}}}\\
{{{\bar x}(0)}}
\end{bmatrix}.$$ Therefore, \eqref{Behavior.proof1} can be transformed into 
\begin{align*}
\begin{bmatrix}
	{H_L^{cum}(\{{u_{i}^{[0,{T_0-1}]}\}_{i=1}^p})}\\
	{H_L^{cum}(\{{x_{i}^{[0,{T_0-1}]}\}_{i=1}^p})}
\end{bmatrix}g &= \begin{bmatrix}
	I&0\\
	{{{\mathcal T}_L}}&{{{\mathcal O}_L}}
\end{bmatrix}\begin{bmatrix}
	{{{\bar u}^{[0,L - 1]}}}\\
	{{{\bar x}(0)}}
\end{bmatrix}     \\
&= \begin{bmatrix}
	{{{\bar u}^{[0,L - 1]}}}\\
	{{{\bar x}^{[0,L - 1]}}}
\end{bmatrix},
\end{align*}
which leads to conclusion 1). The proofs of conclusions 2) and 3) follow a similar process and are omitted here.

\hfill $\blacksquare$

Similar to the original fundamental lemma, Lemma \ref{lem.extend} establishes that \(L\)-long trajectories can be expressed as linear combinations of columns of certain matrices. It maintains reliance on excitation conditions (CCPE, MCPE, HCPE) to ensure the validity of the representation. However, this lemma explicitly addresses \(p\)-segment input-state trajectories, relaxes the PE condition, and allows the excitation condition to hold collectively across all segments rather than individually for each segment. 

By introducing CCPE, MCPE, and HCPE, the extended lemma formalizes the conditions under which segmented data collectively provide sufficient information for trajectory representation. This generalization broadens the lemma’s applicability to scenarios such as systems with distributed sensing, segmented experiments, or partial data. Notably, the work in  \cite{9062331} marks an initial step toward addressing multiple trajectories in Willems' Lemma, while Lemma \ref{lem.extend} strictly extends and incorporates the results from  \cite{9062331}. In special or complex cases (e.g., significant numerical discrepancies between trajectories or equal trajectory lengths), the additional flexibility afforded by Lemma \ref{lem.extend} enhances its applicability.

\begin{remark}  
Willems' fundamental lemma has become a cornerstone in the development of data-driven control methodologies. Lemmas \ref{lem.CPE.Rank} and \ref{lem.extend} share core properties with Willems' lemma, positioning them as fundamental tools for similar applications. Specifically, Lemma \ref{lem.CPE.Rank} enables results analogous to \cite[Th. 3]{8933093}, where the problem of stabilizing controller design is reformulated as a linear matrix inequality (LMI) optimization task. Additionally, Lemma \ref{lem.extend} supports the construction of predictive models, providing a foundation for data-driven model predictive control (MPC) designs, as demonstrated in works such as \cite{9109670,9744574,9903320}. Therefore, Lemmas \ref{lem.CPE.Rank} and \ref{lem.extend} extend the applicability of data-driven techniques to scenarios involving multiple a priori trajectories, offering broader prospects for practical applications. Moreover, while the identification methods and extension lemma presented in this paper are based on fully observable LTI systems, they can be easily extended to input-output state-space systems, as the core application of these methods is rooted in Lemma \ref{lem.CPE.Rank} , which holds for input-output systems as well.
\end{remark}  

\begin{remark}
Both applications discussed above rely on pre-collected data and employ non-adaptive methods. In reality, all three informativeness conditions can be reformulated to resemble the common assumptions about signals in adaptive methods. Specifically, since \( H_L^{mos}(\{z_i^{[0,T_i-1]}\}_{i=1}^{p}) \) has full row rank, it follows that
$$H_L^{mos}(\{z_i^{[0,T_i-1]}\}_{i=1}^{p}) \left(H_L^{mos}(\{z_i^{[0,T_i-1]}\}_{i=1}^{p})\right)^\top > 0.$$

Therefore, there exists a constant \( \beta_0 > 0 \) such that
$$\sum_{i=1}^{p} \sum_{j=0}^{L-1} \alpha_i^2 \phi_{i,j} \phi_{i,j}^\top > \beta_0 I,$$
where $\phi_{i,j} = \left[ z_i(j) \quad z_i(j+1) \quad \cdots \quad z_i(j+T_i-L) \right].$

This condition is analogous to the informativeness condition in adaptive methods. According to Theorem \ref{lem.transformative}, the same holds for both the CCPE and HCPE conditions. Consequently, the informativeness condition defined in this paper has significant potential for application to adaptive data-driven methods as well. Detailed results regarding the three conditions in distributed adaptive system identification methods are elaborated in the extended version \citep{cao2025informativityconditionsmultiplesignals}.
\end{remark}

\section{Experimental Design for the CPE Conditions}\label{s4}
In this section, we propose the experimental design approach to fulfill the three CPE conditions. This methodology is motivated by two critical challenges in practical applications. 1) Overcoming the limitation of single-trajectory PE requirements: Traditional PE conditions demand that individual trajectories exhibit sufficient richness (i.e., full rank conditions), which can be restrictive or infeasible in real-world scenarios (e.g., sparse or intermittent data). By contrast, our approach relaxes this stringent requirement, allowing non-PE signals to collectively achieve CPE through strategic design. 2) Unified online/offline applicability: Many existing methods are limited to either offline (pre-recorded data) or online (real-time) settings, but not both. Our framework is deliberately designed for dual-mode compatibility, ensuring flexibility in deployment.

\textbf{\textit{Signal design for the MCPE condition:} }
For the MCPE condition, the design ensures that the matrix ${H_L^{mos}(\{z^{[0,T_i-1]}_i\}_{i=1}^{p})}$ takes the following form:
\begin{figure}[htbp]
\centering
\includegraphics[scale=1.5]{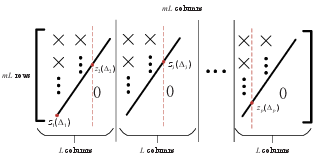}
\label{MCPE_condition}
\end{figure}

Each group of $L$ columns forms an approximately upper triangular structure, with the diagonal elements of the first $m$ submatrices being linearly independent. To construct the desired matrix using $p$ signals, it is necessary to identify which of the first $L$ values of each signal appears as diagonal elements in the submatrices, i.e., $z_i(\Delta(i))$, where $\Delta_{1}=L-1$, $\Delta_{i}  = L-1-\text{mod}(\sum^{i-1}_{j=1} (T_{j}-L+1),L)$ for $i>1$. 

If \(T_{i-1} \geq \Delta_{i-1} + L\), then a submatrix spans from $z_{i-1}(\Delta(i-1))$ to $z_i(\Delta(i))$, and $z_i(\Delta(i))$ must be designed such that $z_{i}(\Delta_{i}) \notin \text{im}\ [z_{1}(\Delta_{1}) \cdots z_{i-1}(\Delta_{i-1})]$. Otherwise, the submatrix is not spanned, and \(z_i(\Delta(i))\) is chosen as  \(z_{i}(\Delta_{i}) \in \text{im}\ z_{i-1}(\Delta_{i-1})\). 

For \(k > L-1\), if \(z_i(k)\) corresponds to the start of the diagonal of a submatrix, i.e.,  \(T_i^{s}+k \in \{2L-1,3L-1,\ldots,(m+1)L-1\}\), where $T_1^{s}=0$, and $T_i^{s}=\sum^{i-1}_{j=1} (T_{j}-L+1)$ for $i>1$, it is necessary to verify whether \(z_i\) can still form \(L\) columns. If feasible, \(z_i(k)\) is designed to remain linearly independent from the existing diagonal elements. If this is not feasible, then $z_{i}(k)$ need to satify \(z_{i}(k) \in \text{im}\  z_{i+1}(\Delta_{i+1})\). The detailed design process is shown below: 

Consider multiple signals $\{z^{[0,T_i-1]}_i\}_{i=1}^{p}$, where $z_{i}(k)\in \mathbb{R}^{m} $ for $k= 0, 1, ... ,{T_i-1}$. Without loss of generality, assume that $L\leq T_1 \leq T_2\leq...\leq T_p<(m+1)L-1$. 

Construction of $z_i(0), z_1(1), \ldots, z_i(L-1)$ for $i=1,\ldots, p$:
\begin{itemize}
\item	
For $i=1$:

Select $z_i^{[0,L-2]}$ arbitrarily, and set $\text{rank}\  z_i(L-1)=1$.
\end{itemize}
\begin{itemize}
\item	
For $i>1$:

Select $z_i^{[0,\Delta_{i} -1]}$ arbitrarily for $\Delta_{i}>1$, and choose $z_i(\Delta_{i})$ as follows: 

$\begin{cases}z_{i}(\Delta_{i})\notin \text{im}\  [z_{1}\left( \Delta_{1} \right)  \  \cdots \  z_{i-1}(\Delta_{i-1} )],&\\
	\  \  \  \ \ \ \text{if}\ T_{i-1}\geqslant \Delta_{i-1}+L\ \text{and} \  T_i^s<\left( m+1\right)  L-1, &\\
	z_{i}\left( \Delta_{i} \right)  \in \text{im}\ z_{i-1}\left( \Delta_{i-1} \right) ,\ \ \ \text{otherwise}.& \end{cases} $

Then, set $\ z^{[\Delta_{i} +1,L-1]}_{i}=0$ for $ \Delta_{i}<L-1$.
\end{itemize}

Construction of $z_i(L), z_i(L+1), \ldots, z_i(T_i-1)$ for $i=1,\ldots, p$:
\begin{itemize}
\item	
For $k<T_i$ and $T_i^{s}+k\in\{2L-1,3L-1,\ldots,(m+1)L-1\}$:

Select $z_i(k)$ as follows:

$\begin{cases}z_{i}(k)\notin\text{im}\  [z_{1}\left( \Delta_{1} \right)  \cdots z_{p}(\Delta_{p} )\ \bar{z}_1  \cdots \bar{z}_p],&\text{if} \  T_{i}-k\geqslant L,\\ z_{i}\left( k\right)  \in \text{im}\ z_{i+1}\left( \Delta_{i+1} \right),  &\text{otherwise} ,\end{cases}  $
where $\bar{z}_i= [z_i(L-1)\  z_i(2L-1) \cdots]$.
\end{itemize}
\begin{itemize}
\item	
For $k<T_i$ and $T_i^{s}+k\notin\{2L-1,3L-1,\ldots,(m+1)L-1\}$:

Set $z_i(k)=0$.
\end{itemize}

\textbf{\textit{Signal design for the CCPE condition:} }The idea behind the design of the CCPE condition is also to construct the first $mL$ columns of ${H_L^{cum}(\{z^{[0,T_0-1]}_i\}_{i=1}^{p})}$ to be a matrix similar to the previous one. The detailed design process is shown below: 

Consider multiple signals $\{z^{[0,T_0-1]}_i\}_{i=1}^{p}$, where $z_{i}(k)\in \mathbb{R}^{m}$ for $k= 0, 1, ... ,{T_0-1}$. Define $Z_i^j = \left[ z_{i}\begin{matrix}\left( j\right)  &z_{i}\left( j+L\right)  &\cdots &z_{i}(j+ \left( m-1)L\right)  \end{matrix} \right]$.

Construction of $z_i(0), ..., z_i(mL-1)$ for $i=1,\ldots, p$:
\begin{itemize}
\item 
Choose $z_i(k)$ such that $\text{rank}\ \sum^{p}_{i=1} \alpha_i z_{i}\left( k\right)=1$ for $k\in\{L-1,2L-1,\ldots,mL-1\}$, and $\sum^{p}_{i=1} \alpha_iz_{i}\left( k\right)  =0$ otherwise.

\end{itemize}
\begin{itemize}
\item 
Ensure that 
\begin{align*}
	\sum^{p}_{i=1} \alpha_{i} \left[ z_{i}\begin{matrix}\left( L-1\right)& z_{i}\left( 2L-1\right)  &\cdots &z_{i}\left( mL-1\right)  \end{matrix} \right] 
\end{align*}
has full rank.
\end{itemize}
\begin{itemize}
\item Ensure $\text{rank} \ Z_i^{0}=m$  for $i=1,\ldots, p$.
\end{itemize}

Construction of $z_i(k), k>mL-1$ for $i=1,\ldots, p$:
\begin{itemize}
\item  Design $z_{i}\left( k\right)$ as
\begin{align}\label{Construction of CCPE}
	z_{i}\left( k\right)  =Z^{(k-mL+1)}_{i}({Z^{0}_{i})}^{-1}  z_{i}\left( mL-1\right).
\end{align}
\end{itemize}

\textbf{\textit{Signal design for the HCPE condition:} } The design idea for the HCPE condition combines the MCPE and CCPE conditions as shown below:

Consider multiple signals $\{z^{[0,T_0-1]}_i\}_{i=1}^{\bar{p}}$ and $\{z^{[0,T_i-1]}_i\}_{i=\bar{p}+1}^{p}$, where $z_{i}(k)\in \mathbb{R}^{m}$, $k= 0, 1, \ldots, {T_i-1}$. Without loss of generality, assume that $L\leq T_{\bar{p}+1} \leq T_{\bar{p}+2}\leq...\leq T_p<(m+1)L$. 

Construction of $z_i(0), \ldots, z_i(T_0-1)$ for $i=1,\ldots, \bar{p}:$
\begin{itemize}
\item 
If $T_0\geqslant (m+1)L$:

Use the CCPE signal design method.
\end{itemize}
\begin{itemize}
\item 
If $T_0< (m+1)L$:

Choose $z_i(k)$ such that $\text{rank}\ \sum^{\bar{p}}_{i=1} \alpha_i z_{i}\left( k\right)=1$ for $k\in\{L-1,2L-1,\ldots,mL-1\}$, and $\sum^{\bar{p}}_{i=1} \alpha_iz_{i}\left( k\right)  =0$ otherwise. Ensure that 
\begin{align*}
	\sum^{\bar{p}}_{i=1} \alpha_{i} \left[ z_{i}\begin{matrix}\left( L-1\right)& z_{i}\left( 2L-1\right)  &\cdots \end{matrix} \right] 
\end{align*}
has full column rank.
\end{itemize}

Construction of $z_i(0),\ldots, z_i(T_i-1)$ for $i=\bar{p}+1,\ldots, p$:
\begin{itemize}
\item
Let $\sum^{\bar{p}}_{i=1} \alpha_{i}z_i$ represent the initial signal, with $\{z_i\}_{i=\bar{p}+1}^{p}$ serving as the subsequent signals. Employ the MCPE signal design method to construct $z_i(0), \ldots, z_i(T_i-1)$ for $i=\bar{p}+1, \ldots, p$.
\end{itemize}

\begin{theorem}\label{Proof of experimental design}
For multiple signals $\{z^{[0,T_i-1]}_i\}_{i=1}^{p}$, where $z_{i}(k)\in \mathbb{R}^{m}$ for $k= 0, 1, \ldots,{T_i-1}$. 
\begin{itemize}
	\item[$1).$] 
	Using the MCPE design method, the signal length required to satisfy $\text{rank}\  {H_L^{mos}(\{z^{[0,T_i-1]}_i\}_{i=1}^{p})}=mL$ is $\sum\limits_{i = 1}^p {{T_i}} \ge mL + p(L - 1)$. 
\end{itemize}
\begin{itemize}
	\item[$2).$] 
	Using the CCPE design method, the signal length required to satisfy $\text{rank}\  {H_L^{cum}(\{z^{[0,T_0-1]}_i\}_{i=1}^{p})}=mL$ is ${T_0} \ge (m+1)L  - 1$. 
\end{itemize}
\begin{itemize}
	\item[$3).$] 
	Using the HCPE design method, the signal length required to satisfy $\text{rank}\  {H_L^{hyb}(\{z^{[0,T_i-1]}_i\}_{i=1}^{p})}=mL$ is $T_0 + \sum\limits_{i = \bar{p}}^p {{T_i}} \ge mL + (p-\bar{p}+1)(L - 1)$. 
\end{itemize}

Additionally, all three design methods ensure that each sub-signal is non-persistently exciting, i.e., $\text{rank}\ {H_L}({z_{i}^{[0, T_i - 1]}}) < mL$.
\end{theorem}
\textbf {Proof.}  1). The design choices ensure that the \(p\) signals collectively construct the desired matrix while satisfying the length constraint \(T_i < (m+1)L\) if \(p > 1\). Therefore, 
\[
\text{rank}\ {H_L}({z_{i}^{[0, T_i - 1]}}) < mL.
\]

2). Since $\text{rank}\ \sum^{p}_{i=1} \alpha_i z_{i}\left( k\right)=1$ for $k\in\{L-1,2L-1,....,mL-1\}$, and $\sum^{p}_{i=1} \alpha_iz_{i}\left( k\right)  =0$ otherwise, we have 
$$\text{rank}\  {H_L^{cum}(\{z^{[kL,(k+2)(L-1)]}_i\}_{i=1}^{p})}=L$$
for all $ k\text{} \in \left\{ 0,1,...,m-1\right\} $. At this point, $ {H_L^{cum}(\{z^{[kL,(k+2)(L-1)]}_i\}_{i=1}^{p})}$ takes the form of a lower triangular matrix with nonzero diagonal elements, i.e., 
\begin{align*}
&{H_L^{cum}(\{z^{[kL,(k+2)(L-1)]}_i\}_{i=1}^{p})}\\
&=\begin{bmatrix}
	0	& & &z^{cum}_{(k+1)(L-1)} \\
	&          & \iddots & z^{cum}_{k(L-1)+L}  \\
	&z^{cum}_{(k+1)(L-1)}  &\iddots & \vdots \\
	z^{cum}_{(k+1)(L-1)}      & z^{cum}_{k(L-1)+L}            & \cdots & z^{cum}_{(k+2)(L-1)}
\end{bmatrix},
\end{align*}
where 	$z^{cum}_{k}=\sum^{p}_{i=1} \alpha_iz_{i}\left( k\right) $.

Since	$\sum^{p}_{i=1} \alpha_{i} \left[ z_{i}\begin{matrix}\left( L-1\right)& z_{i}\left( 2L-1\right)  &\cdots &z_{i}\left( mL-1\right)  \end{matrix} \right] $ has full rank, the choice of $z_i(k)$, $k\notin\{L-1,2L-1,....,mL-1\}$ at each time step increases the rank of the cumulative Hankel matrix ${H_L^{cum}(\{z^{[0,T_0-1]}_i\}_{i=1}^{p})}$ until it becomes full rank at $T_0=(m+1)L$. 

Next, we demonstrate that designing $z_i(k)$ for $k >m L-1$ according to the proposed approach ensures that each sub-Hankel matrix does not satisfy $mL$-order persistently exciting. Specifically, since $Z_i^{0}$ is full rank for $i = 1, \dots, p$, there always exists a design law as described in  \eqref{Construction of CCPE}. Consequently, we have
$$z^{[k,k+L-1]}_{i}\in \text{im} \left[ \begin{matrix}Z^{k-mL+1}_{i}\\ Z^{k-mL+2}_{i}\\ \vdots \\ Z^{k-(m-1)L}_{i}\end{matrix} \right]   , k\geqslant mL-1$$
for all $i \in \{1, \dots, p\}$. This result implies that the rank of the Hankel matrix $H_L(z_i^{[0,T_0-1]})$ cannot increase beyond the $m L$-th column. Therefore,
$$\text{rank}\ {H_L}({z_{i}^{[0,{T_0} - 1]}})<mL.$$

3). Building upon the proof procedures for signal design under the MCPE and CCPE conditions, the HCPE condition can be similarly achieved. Specifically, the matrix  $H_L^{hyb}(\{z^{[0,T_i-1]}i\}{i=1}^{p})$  attains full row rank if  $T_0 + \sum_{j=\bar{p}+1}^{p} T_j \geq (m+1)L$,  and each signal exhibits non-$mL$-order persistently exciting. \hfill $\blacksquare$

\begin{remark}
Recent finite-time experiment design methods (e.g., \cite{10601334}) optimize a single exploration trajectory, typically harmonic or multisine inputs, to minimize parameter uncertainty and guarantee a priori performance bounds after identification. In contrast, the goal of our experiment design is not to minimize estimation error for a pre-specified model, but rather to ensure that the data collected across possibly multiple trials satisfy the proposed CPE conditions, which are sufficient for data-driven controller synthesis. This is crucial in settings where a single long PE trajectory cannot be executed due to safety or operational constraints. Moreover, our approach improves the numerical conditioning of the resulting Hankel matrix via weighting, thereby reducing the number of experimental repetitions needed to achieve data informativity. Hence, while targeted exploration focuses on minimizing model uncertainty for robust performance, our design explicitly guarantees the feasibility of direct data-driven control synthesis from multiple finite trajectories, even under limited excitation budgets.
\end{remark}

\begin{remark}
According to Theorem \ref{Proof of experimental design}, all three conditions can be satisfied within the minimum number of time steps. Specifically, if we fix ${H_L^{mos}(\{z^{[0,T_i-1]}_i\}_{i=1}^{p})}$, ${H_L^{cum}(\{z^{[0,T_0-1]}_i\}_{i=1}^{p})}$, and ${H_L^{hyb}(\{z^{[0,T_i-1]}_i\}_{i=1}^{p})}$ as standard square matrices, the experimental design method outlined in Section \ref{s4} can achieve the required conditions. In this scenario, the eigenvalues of the three composite matrices depend solely on the elements along the diagonal of each $L$ column. Furthermore, we note that the experimental design methods for all three conditions do not require knowledge of future information, allowing them to be implemented either offline or online. This enables real-time adjustment of control signals based on system state constraints while satisfying design requirements, thereby preventing extreme state divergence.
\end{remark}

\section{ Application in Distributed Adaptive Identification}
In this section, we explore the shift from single trajectories for identification to distributed identification using multiple trajectories that satisfy any CPE condition. Both single and distributed identifiers, our approach is direct, online and adaptive. Additionally, we highlight that the devised identification technique remains unaffected by the signal's upper limit, making it highly robust as well.

Before proceeding, we establish exponential stability conditions for the autonomous system 
\begin{align} \label{Excited autonomous system}
	x(k+1) = (I_n-z(k)z^\top(k))x(k),
\end{align}
where $x(k)\in \mathbb{R}^n$, $z(k)\in\mathbb{R}^{n\times m}$, at the equilibrium point $x_e = 0$. This system is a simplified error expression for the subsequent single-trajectory based identifier and multi-trajectory based distributed identifier. Moreover, it forms the theoretical foundation for analyzing the exponential convergence properties of the identifiers.

\begin{lemma}\label{lem.exponential.stability}
	For the system \eqref{Excited autonomous system}, the equilibrium state $x_e=0$ is exponentially stable (e.s.) if the following conditions are satisfied for all $k\geq0$:
	\begin{itemize}
		\item[\romannumeral1).] 
		The signal $z(k)$ is bounded, i.e. there exists a constant $z_m>0$ such that $||z(k)|| < {z_m}$;
	\end{itemize}
	\begin{itemize}
		\item[\romannumeral2).] 
		$2I - {z^\top}(k)z(k) > 0$;
	\end{itemize}
	\begin{itemize}
		\item[\romannumeral3).] 
		$z^{[k,k+l-1]}$ is PE of order $1$ for a fixed constant $l\geq1$.
	\end{itemize}
\end{lemma}

{\it Proof:} See the Appendix. \hfill $\blacksquare$

\subsection{Adaptive Identifier for a Single Control Sequence Satisfying PE Condition}
In this subsection, we devise an adaptive identifier for a single trajectory sequence within the framework established by Lemma \ref{lem.exponential.stability}, using the control sequence necessitating persistent excitation. In the subsequent subsections, we extend the results to  distributed adaptive identifiers for multi-trajectory sequences, where each of the control sequence can come from either smooth or low-rank controller without having persistently exciting.

To facilitate system model identification, we transform the LTI system \eqref{Sysform. (1)} into 
\begin{align} \label{Sysform. (2)}
	x(k+1) = r^\top(k)\theta,
\end{align}
where $r(k) = I_n \otimes \begin{bmatrix}
	{x(k)}\\
	{u(k)}
\end{bmatrix}\in {\mathbb{R}^{n(n+m) \times {n}}}$, $\theta = \begin{bmatrix}
	A &B
\end{bmatrix}_{vec}\in {\mathbb{R}^{n(n+m)}}$. 

For systems represented by \eqref{Sysform. (2)}, a discrete-time adaptive identifier is designed as
\begin{align} \label{Ideform. (1)}
	\theta (k + 1) = \theta (k) + \alpha r(k){\left( {{r^\top}(k)r(k) + \xi I} \right)^{ - 1}}\varepsilon (k),
\end{align} 
where $\theta (k)$ is the identification vector at moment $k$, $\alpha >0$ denotes the adaptive gain, $\xi>0$ is an arbitrary positive constant, and $\varepsilon (k)={x(k + 1) - {r^\top}(k)\theta (k)}$. 

Then, the dynamics of the identification error $\tilde \theta (k) = \theta (k) - \theta $ of \eqref{Ideform. (1)} can be expressed as 
\begin{align}\label{Ide_error}
	\tilde \theta (k + 1) = (I - \alpha r(k){\left( {{r^\top}(k)r(k) + \xi I} \right)^{ - 1}}{r^\top}(k))\tilde \theta (k).
\end{align}

Since ${{r^\top}(k)r(k) + \xi I}$ is an invertible symmetric matrix, \eqref{Ide_error} possesses the same form as \eqref{Excited autonomous system}. We can use Lemma \ref{lem.exponential.stability} to establish the exponential convergence condition of the identifier \eqref{Ideform. (1)}, as shown in the following theorem.
\begin{theorem}\label{The.single}
	For the identifier \eqref{Ideform. (1)}, $\theta (k)\rightarrow \theta$ exponentially fast if the following  conditions hold for all $k\geq 0$:	
	\begin{itemize}
		\item[\romannumeral1).] 
		$r(k)$ is bounded;
	\end{itemize}
	\begin{itemize}
		\item[\romannumeral2).] 
		$0 < \alpha \leq 2$;
	\end{itemize}
	\begin{itemize}
		\item[\romannumeral3).] 
		$r^{[k,k+l-1]}$ satisfies PE condition of order $1$ with a fixed constant $l\geq1$.
	\end{itemize}
\end{theorem}

{\it Proof:} 
Utilizing the Woodbury Matrix Identity, we obtain  $r{({r^\top}r + \xi I)^{ - 1}}{r^\top} = I - {(I + \frac{1}{\zeta }r{r^\top})^{ - 1}} < I$. Consequently,
\begin{align}\label{WMI}
	2I - \alpha r(k){\left( {{r^\top}(k)r(k) + \xi I} \right)^{ - 1}}{r^\top}(k)>0
\end{align}
when $0<\alpha\leq2$, thereby fulfilling condition \romannumeral2) of Lemma \ref{lem.exponential.stability}. The remaining assumptions in Theorem \ref{The.single} correspond to conditions \romannumeral1) and \romannumeral3) in Lemma \ref{lem.exponential.stability}, respectively. Hence, $\tilde \theta = 0$ is exponentially stable, implying that $\theta (k)$ converges exponentially fast to $\theta$.
\hfill $\blacksquare$

Note that for the systems described by \eqref{Sysform. (1)} or \eqref{Sysform. (2)}, when a set of input-state data can uniquely determine $A$ and $B$, it is typically necessary for the input sequence to constitute a persistently exciting sequence of order $n + 1$.  If the excitation condition for $u_{[k,k+l-1]}$ holds for all $k\geq0$ with a fixed constant $l\geq1$, then $\begin{bmatrix}
	{{H_1}({x^{[k,k+l-1]}})}\\
	{{H_1}({u^{[k,k+l-1]}})}
\end{bmatrix}$ achieves full row rank. Consequently, ${H_1}({r^{[k,k+l-1]}})$ also attains full row rank, meeting the excitation requirement outlined in Theorem \ref{The.single}. Therefore, the following corollary ensues.
\begin{corollary}\label{corollary to single identifier}
	For the identifier \eqref{Ideform. (1)}, $\theta (k)\rightarrow \theta$ exponentially fast if the following  conditions hold for all $k\geq 0$:	
	\begin{itemize}
		\item[\romannumeral1).] 
		$x(k)$ and $u(k)$ are bounded;
	\end{itemize}
	\begin{itemize}
		\item[\romannumeral2).] 
		$0 < \alpha \leq 2$;
	\end{itemize}
	\begin{itemize}
		\item[\romannumeral3).] 
		$u^{[k,k+l-1]}$ satisfies PE condition of order $n+1$ with a fixed constant $l\geq1$.
	\end{itemize}
\end{corollary}

\subsection{Distributed Adaptive Identifiers for Multiple Control Sequences Satisfying CPE Conditions}

In this subsection, we will leverage the CPE condition to formulate distributed adaptive identifiers with smooth input signals within the state-space setting, elucidating their conditions for exponential stability. Each sequence of trajectories used need not satisfy the identifiability condition, and can also enables the synthesis of control inputs for the desired system through the stabilizing controllers of multiple subsystems without necessitating perturbation experiments.

Consider a  homogeneous multi-agent system (MAS) with $N$ agents, the dynamics of each agent are given by
\begin{align} \label{Sysform. (3)}
	x_i(k+1) = r_i^\top(k)\theta, \ \ i=1,2,...,N
\end{align}
where $r_i(k) = I \otimes \begin{bmatrix}
	{x_i(k)}\\
	{u_i(k)}
\end{bmatrix}\in {\mathbb{R}^{n(n+m) \times {n}}}$, $\theta = \begin{bmatrix}
	A &B
\end{bmatrix}_{vec}\in {\mathbb{R}^{n(n+m)}}$.

Based on the previous identifier \eqref{Ideform. (1)} for a single trajectory, the distributed identifier for the system \eqref{Sysform. (3)} is designed as
\begin{align}\label{Ideform. (2)}
	{\theta _i}(k + 1) = {\theta _i}(k) + &\alpha {r_i}(k){\left( {{r_i}^\top(k){r_i}(k) + \xi I} \right)^{ - 1}}{\varepsilon _i}(k)\nonumber\\
	&~~~~~~- \gamma \sum\limits_{j \in {\mathcal{N}_i}} {\left( {{\theta _i}(k) - {\theta _j}(k)} \right)}, 
\end{align}
where $\theta_i (k)$ is the identification vector of $i$-th agent at moment $k$, $\alpha>0$ and $\gamma>0$ are constants to be designed, $\xi>0$ is an arbitrary positive constant, and $\varepsilon_i (k)={x_i(k + 1) - {r_i^\top}(k)\theta_i (k)}$.

Let ${{\tilde \theta }_i}(k) = {\theta _i}(k) - \theta$ and ${\tilde\theta _c}(k)= \begin{bmatrix}
	{\tilde\theta _1^\top(k)},{\tilde\theta _2^\top(k)},...,{\tilde\theta _N^\top(k)}
\end{bmatrix}^\top$ denote the identification error of the $i$-th identifier and the compact form of the errors, respectively. Then, the error dynamics can be expressed as
\begin{align}\label{compactError}
	\tilde \theta_c (k + 1) = \left( {I - \alpha R(k) - \gamma {\mathcal L} \otimes I} \right)\tilde \theta_c (k), 
\end{align}
where $R(k)=\text{diag}(\bar r_1(k),...,\bar r_N(k))$ and $\bar r_i(k)= r_i(k){\left( {{r_i^\top}(k)r_i(k) + \xi I} \right)^{ - 1}}{r_i^\top}(k)$.

\begin{theorem}\label{The.multi}
	Suppose that the communication network $\mathcal{G}$ is undirected. Then, $\theta_i (k)\rightarrow \theta$, $i=1,2,...,N$ exponentially fast if the following  conditions hold for all $k\geq 0$:	
\end{theorem}
\begin{itemize}
	\item[\romannumeral1).] 
	The input-state data of the MAS is bounded;
	\item[\romannumeral2).] 
	$0<\alpha\leq2$ and $0<\gamma<\frac{1}{{{\lambda _{\max }}(\mathcal{L})}}$;
	\item[\romannumeral3).] 
	The inputs $\{u_i^{[k,{k + l-1}]}\}_{i=1}^N$ are any type of CPE of order $n+1$ for a fixed constant $l\geq1$.
\end{itemize}

{\it Proof:} By definition, we know that $\alpha R(k) +\gamma {\mathcal L} \otimes I$ is a semi-positive definite matrix, so there exists a matrix $\Phi(k)$ such that $\Phi(k)\Phi^\top(k) = \alpha R(k) +\gamma {\mathcal L} \otimes I$. Then, \eqref{compactError} can be written as
\begin{align}
	\tilde \theta_c (k + 1) = \left( {I - \Phi(k)\Phi^\top(k)} \right)\tilde \theta_c (k). 
\end{align}

We first prove that $\Phi^{[k,k+l-1]}$ is PE of order $1$ for a fixed constant $l\geq1$ when $\{u_i^{[k,{k + l-1}]}\}_{i=1}^N$ are MCPE of order $n+1$. The problem can be translated into proving that
\begin{align*}
	\sum\limits_{t = 0}^{l - 1} {\Phi (k + t){\Phi ^\top}(k + t)}  = l\gamma \mathcal{L} \otimes I + \alpha\sum\limits_{t = 0}^{l - 1} { R(k + t)}  > 0. 
\end{align*}

The preceding inequality can be equivalently expressed as $\ker \mathcal{L} \otimes I \cap \ker \sum\limits_{t = 0}^{l - 1} { R(k + t)}=\{0\}$. In graph theory, for an undirected and connected communication network $\mathcal{G}$, $\mathcal{L}\otimes I$ possesses $(n+m)n$ zero eigenvalues. The elements in $\ker \mathcal{L} \otimes I$ can be represented as $(\textbf{1}_{(n+m)n}\otimes I_{(n+m)n})\eta$, where $\eta\in\mathbb{R}^{(n+m)n}$. Consequently, we obtain
\begin{align}\label{distributed.proof}
	&{\eta ^\top}{({{\bf{1}}_{(n + m)n}} \otimes {I_{(n + m)n}})^\top}\sum\limits_{t = 0}^{l - 1} {R(k + t)} ({{\bf{1}}_{(n + m)n}} \otimes {I_{(n + m)n}})\eta \nonumber\\
	&~~~~~~~~~~~~~~~~~~~~~~~~~~~= {\eta ^T}\sum\limits_{t = 0}^{l - 1} {\sum\limits_{i = 1}^N {{{\bar r}_i}(k + t)} } \eta.& 
\end{align}

If  $\{u_i^{[k,{k + l-1}]}\}_{i=1}^N$ satisfy the MCPE condition, it can be readily inferred that $\sum\limits_{t = 0}^{l - 1} {\sum\limits_{i = 1}^N {{{\bar r}_i}(k + t)} }>0$. Thus, \eqref{distributed.proof} equals $0$ if and only if $\eta$ is the zero vector, implying that $\ker \mathcal{L} \otimes I \cap \ker \sum\limits_{t = 0}^{l - 1} { R(k + t)}=\{0\}$. Moreover, according to Lemma \ref{lem.transformative}, if $\{u_i^{[k,{k + l-1}]}\}_{i=1}^N$ satisfy the CCPE or HCPE condition, they must also satisfy the MCPE condition. Therefore,  $\Phi^{[k,k+l-1]}$ are PE of order $1$ for all types of CPE of order $n+1$.

Next we prove that $2I - \Phi^\top(k)\Phi(k)>0$. Let $\alpha R(k)=\phi_1(k)\phi_1^\top(k)$ and $\gamma {\mathcal L} \otimes I=\phi_2(k)\phi_2^\top(k)$, thus we have $\Phi(k)=\begin{bmatrix}{\phi_1(k)} & {\phi_2(k)}\end{bmatrix}$. Then, the inequality $2I - \Phi^\top(k)\Phi(k)>0$ is equivalent to 
\begin{align}\label{inequality2}
	\begin{bmatrix}
		{2I - \phi _1^\top(k){\phi _1}(k)}&-{\phi _1^\top(k){\phi _2}(k)}\\
		-{\phi _2^\top(k){\phi _1}(k)}&{2I - \phi _2^\top(k){\phi _2}(k)}
	\end{bmatrix} > 0. 
\end{align}

By \eqref{WMI}, if $0<\alpha\leq2$, then $2I - \phi _1^T(k){\phi _1}(k)>0$. Thus, the Schur complement of \eqref{inequality2} is of the form
\begin{align*}
	2I - \phi _2^\top(k){\phi _2}(k)- \phi _2^\top(k)M_2(k){\phi _2}(k) > 0,
\end{align*}
where $M_2(k)={\phi _1}(k){\left( {2I - \phi _1^\top(k){\phi _1}(k)} \right)^{ - 1}}\phi _1^\top(k)$. According to the Woodbury Matrix Identity, ${M_2}(k) = I - {\left( {I + \frac{1}{2}{\phi _1}(k)\phi _1^\top(k)} \right)^{ - 1}}<I$, which implies that 
\begin{align*}
	2I - \phi _2^\top(k){\phi _2}(k) - \phi _2^\top(k){M_2}(k){\phi _2}(k) > 2I - 2\phi _2^\top(k){\phi _2}(k).
\end{align*}

Hence, \eqref{inequality2} holds if $0<\gamma<\frac{1}{{{\lambda _{\max }}(\mathcal{L})}}$. In conclusion, the conditions given in Theorem \ref{The.multi} ensure that the conditions in Lemma \ref{lem.exponential.stability} hold, so we have $\theta_i (k)\rightarrow \theta$, $i=1,2,...,N$ exponentially fast.
\hfill $\blacksquare$
\begin{remark}
	Both the single identifier \eqref{Ideform. (1)} and the distributed identifiers \eqref{Ideform. (2)} rely solely on specific constants for selecting parameters within their exponentially stabilized convergence conditions. They maintain robustness to trajectory data, imposing no stringent requirements on the exact values of upper bounds, as long as signal trajectories remain bounded. 
\end{remark}

\section{Illustrative Examples of Applications}\label{s6}
\subsection{Distributed Identification from Multiple Control Inputs}

\begin{figure}[htbp]
	\centering
	\includegraphics[scale=1]{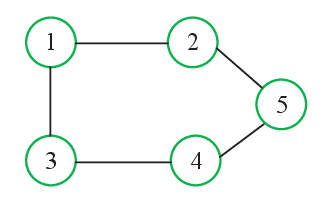}
	\caption {{Network topology composed of all agents.}}
	\label{network}
\end{figure}

In this subsection, we apply the distributed adaptive identification scheme of Section 5 to the voltage converter systems. Consider a homogeneous multi-agent systems comprising five agents, and the communication network $\mathcal{G}$ is shown in Fig. \ref{network}. By discretizing the voltage converter system described in \cite{7409511} with a sampling time of 0.1 seconds, we derive the state-space representation for each agent as
\begin{align*}
	x_i(k+1) = \begin{bmatrix}
		{1.0000}&{ - 0.0500}\\
		{0.0004}&{\ \ 0.9998}
	\end{bmatrix}x_i(k) + \begin{bmatrix}
		{0.0125}\\
		{0.0000}
	\end{bmatrix}u_i(k),
\end{align*}
where $i \in \{1,2,3,4,5\}$. Transforming the system into the form of \eqref{Sysform. (3)} results in $\theta=\begin{bmatrix}
	1.0000&{ - 0.0500}&{0.0125}&{0.0004}&{0.9998}&0.0000
\end{bmatrix}^\top$, where $\theta$ represents the parameters subject to identification.

We choose $\xi = 2$, $\alpha = 1$ and $\gamma = 0.25$, satisfying condition \romannumeral2) in Theorem \ref{The.multi}. A comparison is conducted between the distributed identification scheme delineated in Section 5 and the single-system identification scheme. In the single-system identification scheme, we consider cases with and without random noise. The addition of random noise ensures that each subsystem satisfies the persistently exciting condition in Theorem \ref{The.single} and thus meets the identifiable condition. Conversely, the case without noise is employed to demonstrate the distributed identification scheme's resilience against covert and stealthy attacks.

\begin{figure}[htbp]
	\centering
	\includegraphics[scale=0.6]{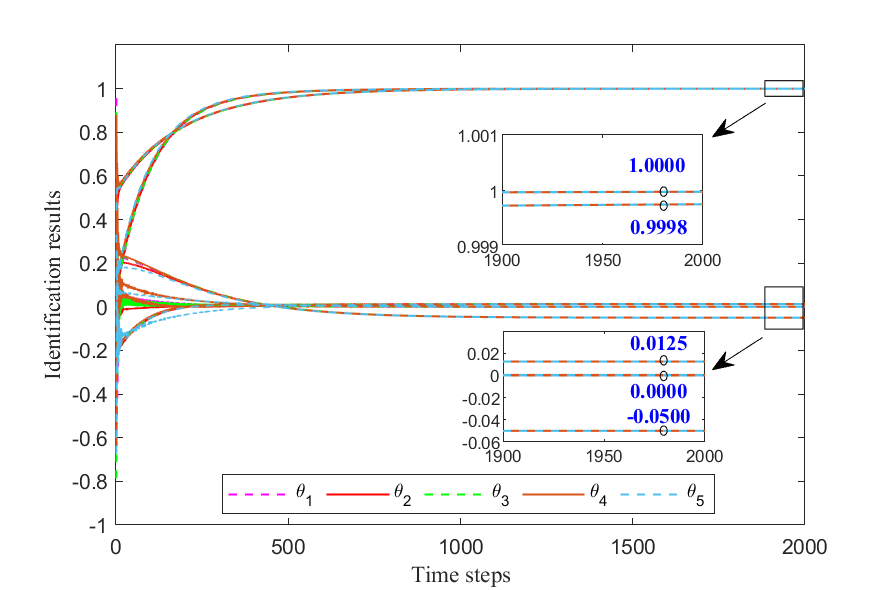}
	\caption {{Distributed adaptive identification from multiple sequences of feedback control inputs.}}
	\label{distributed identification}
\end{figure}

\begin{figure}[htbp]
	\centering
	\includegraphics[scale=0.6]{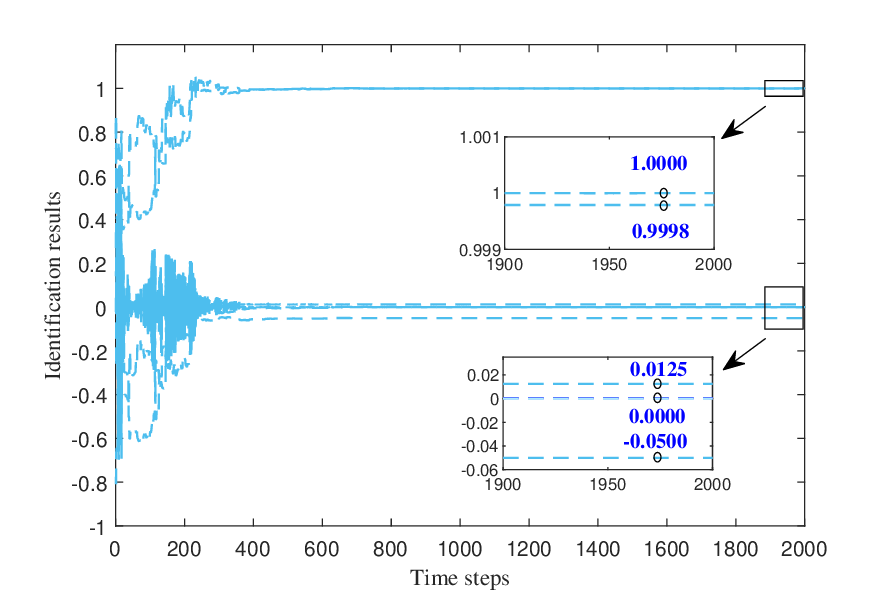}
	\caption {{Single-system identification from a single input sequence with random noise.}}
	\label{single_identification}
\end{figure}

\begin{figure}[htbp]
	\centering
	\includegraphics[scale=0.6]{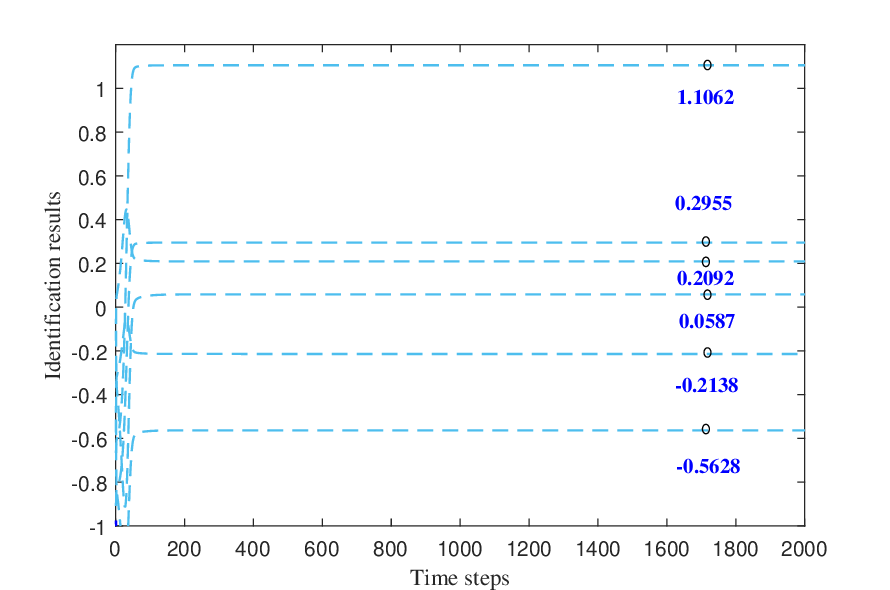}
	\caption {{Single-system identification from a single input sequence without random noise.}}
	\label{single_identification_none}
\end{figure}

Fig. \ref{distributed identification} depicts the distributed adaptive identification results with the collaborative efforts of five agents. It is evident that each agent is able to approximate the desired identification result with a relatively smooth curve when the CPE condition is satisfied. This smoothness arises from the absence of additional noise introduced by the agents' controllers, allowing the multi-agent system to perform the identification through interaction using the smooth system trajectories. 

In contrast, the identification result of a single system, as depicted in Fig. \ref{single_identification}, confirm the identification ability of the identifier \eqref{Ideform. (1)} under persistently exciting condition (see Corollary \ref{corollary to single identifier}). However, due to the requirement for sufficiently stimulating input data, the random noise is injected to the controller, resulting in significant fluctuations during the single-system identification process. The successful identification also indicates a lack of capability to withstand covert, stealth attacks, which is one of the motivations behind our research. 

Fig. \ref{single_identification_none} displays the identification results of a single system without the addition of random noise. It is evident that due to the absence of identifiable conditions, the identifier converges to undesirable results in a short period. This demonstrates the challenge of achieving convergence to the target model using single-input trajectory single-system identification under the influence of bounded feedback controllers without satisfying sustained excitation conditions, further affirming the superiority of the distributed adaptive identification proposed in this paper under CPE conditions. Additionally, it illustrates the effectiveness of distributed identification methods in reducing the risk posed by attacks that require precise modeling.
\subsection{Least-squares Identification from Multiple Control Inputs}\label{s6.A}

In this subsection, we apply the LS identification method to a batch reactor system, previously considered in \cite{8933093}. The system dynamics are described by
\begin{align}\label{batch reactor system}
	x_i(k + 1) &= \begin{bmatrix}
		{1.178}&{0.001}&{0.511}&{ - 0.403}\\
		{ - 0.051}&{0.661}&{ - 0.011}&{0.061}\\
		{0.076}&{0.335}&{0.560}&{0.382}\\
		0&{0.335}&{0.089}&{0.849}
	\end{bmatrix}x_i(k)\nonumber \\
	&~~~~~~~~~~~~~~~~~~~~~	+ \begin{bmatrix}
		{0.004}&{ - 0.087}\\
		{0.467}&{0.001}\\
		{0.213}&{ - 0.235}\\
		{0.213}&{ - 0.016}
	\end{bmatrix}u_i(k)+w_i(k), 
\end{align}
where $w_i(k)\in {\mathbb{R}^n}$ represents the process noise. It is assumed that the process noises is i.i.d.\ Gaussian, i.e., \( w_i(k) \sim \mathcal{N}(0, \sigma_{w}^2 I) \).

This system has a state order of $n = 4$, input order of $m = 2$, and is open-loop unstable.

For each agent, the input/state data is organized as
$$U_i = \left[ \begin{matrix}u_{i}\left( 0\right)  &u_{i}\left( 1\right)  &\cdots &u_{i}\left( T_{i}-1\right)  \end{matrix} \right],  $$
$$X_i^-=\left[ \begin{matrix}x_{i}\left( 0\right)  &x_{i}\left( 1\right)  &\cdots &x_{i}\left( T_{i}-1\right)  \end{matrix} \right] , $$
$$X_i^+=\left[ \begin{matrix}x_{i}\left( 1\right)  &x_{i}\left( 2\right)  &\cdots &x_{i}\left( T_{i}\right)  \end{matrix} \right] , $$
and the process noise data as
$$W_i=\left[ \begin{matrix}w_{i}\left( 0\right)  &w_{i}\left( 1\right)  &\cdots &w_{i}\left( T_{i}-1\right)  \end{matrix} \right],$$
$$W_i^n=\left[ \begin{matrix}w_{i}\left( 0\right)  &w_{i}\left( 1\right)  &\cdots &w_{i}\left( T_{i}-1-n\right)  \end{matrix} \right] . $$

The data combinations in Definitions \ref{CPE-2}-\ref{CPE-3} can be freely selected according to the length of the input data of the $p$ agents. For example, consider the CCPE condition. If $T_1=T_2=...=T_{p}=T_0$, the input data can be combined as $$H^{cum}_{L}\left( \{ U_{i}\}^{p}_{i=1} \right)  =\sum\limits^{p}_{i=1} {{\alpha_{i} }  {H_{L}}  (U_{i})}.$$ 
From the system dynamics \eqref{batch reactor system}, we obtain
\begin{align*}
	&H^{cum}_{1}\left( \{ X^{+}_{i}\}^{p}_{i=1} \right)  \nonumber\\ 
	&=\begin{bmatrix}
		A&B
	\end{bmatrix}\left[ \begin{matrix}H^{cum}_{1}\left( \{ X^{-}_{i}\}^{p}_{i=1} \right)  \\ H^{cum}_{1}\left( \{ U_{i}\}^{p}_{i=1} \right)  \end{matrix} \right] +H^{cum}_{1}\left( \{ W_{i}\}^{p}_{i=1} \right) .
\end{align*}

Define $G=\begin{bmatrix}A&B\end{bmatrix}$. The least squares problem is formulated as
$$\text{} \min_{G} \left\Vert H^{cum}_{1}\left( \{ X^{+}_{i}\}^{p}_{i=1} \right)  -G\left[ \begin{matrix}H^{cum}_{1}\left( \{ X^{-}_{i}\}^{p}_{i=1} \right)  \\ H^{cum}_{1}\left( \{ U_{i}\}^{p}_{i=1} \right)  \end{matrix} \right]  \right\Vert $$

The solution, unique due to Lemma \ref{lem.CPE.Rank}, is given by 
\begin{align}\label{G}
	\hat{G} ={H^{cum}_{1}(\{ X_{i}^+\}^{p}_{i=1} )}  {\begin{bmatrix}{H^{cum}_{1}(\{ X_{i}^-\}^{p}_{i=1} )}  \\ {H^{cum}_{1}(\{ U_{i}\}^{p}_{i=1} )}  \end{bmatrix}^{\dag } } .
\end{align}

In this example, we employ the CCPE, MCPE, and HCPE conditions for LS identification under three different data settings: same-dimensional, different-dimensional, and partially same-dimensional data, respectively. According to \eqref{G}, $\hat{G}$ is identifiable if the matrix $\begin{bmatrix}{H^{cum}_{1}(\{ X_{i}^-\}^{p}_{i=1} )}  \\ {H^{cum}_{1}(\{ U_{i}\}^{p}_{i=1} )}  \end{bmatrix}$ has full row rank. Lemma \ref{lem.CPE.Rank} indicates that this requirement holds as long as the order of the CPE satisfies $L\geq n+1$. For systems with order $n=4$, this implies that the minimum required order of the CPE is $5$. The choice of $L$ involves a trade-off between data length, numerical conditioning, and computational cost. A larger $L$ can improve subspace separation but requires longer experiments and heavier computation, and may worsen conditioning under noise. In practice, it is recommended to select the smallest $L$ that meets the theoretical rank condition while ensuring acceptable numerical properties and manageable experiment duration. To synthesize the necessary CPE conditions of order $5$, we utilize 10 independently input trajectories ($p= 10$), each of length $T_i$ samples.
\begin{figure}[htbp]
	\centering
	\includegraphics[scale=0.4]{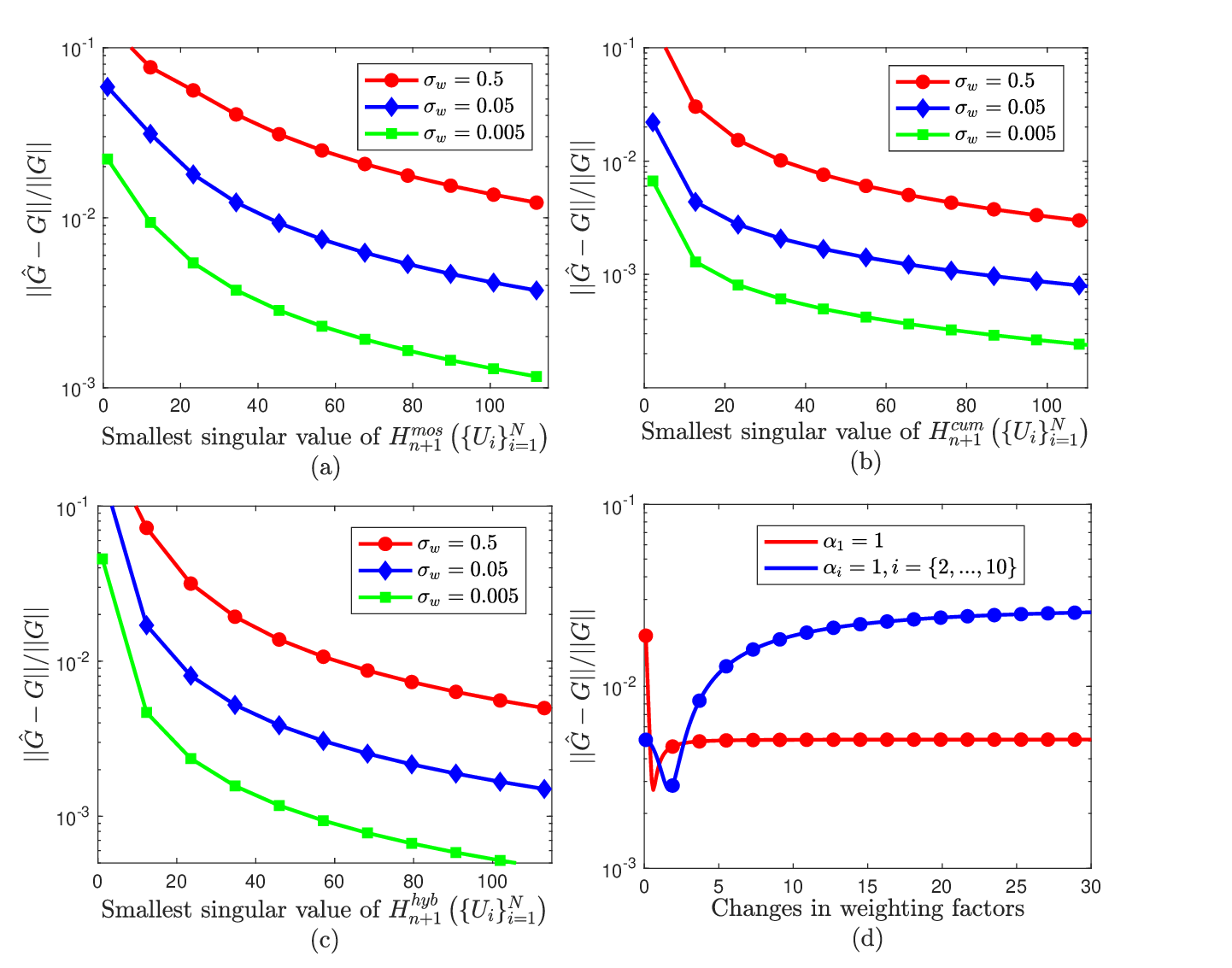}
	\caption {{Identification errors under different settings. (a)-(c) Identification results using the MCPE, CCPE, and HCPE conditions, respectively, across different noise levels and synthetic matrix singular value settings. (d) Identification results for the MCPE condition under various weighting factor configurations.}}
	\label{LS}
\end{figure}

\begin{figure*}[htbp]
	\centering
	\includegraphics[scale=0.60]{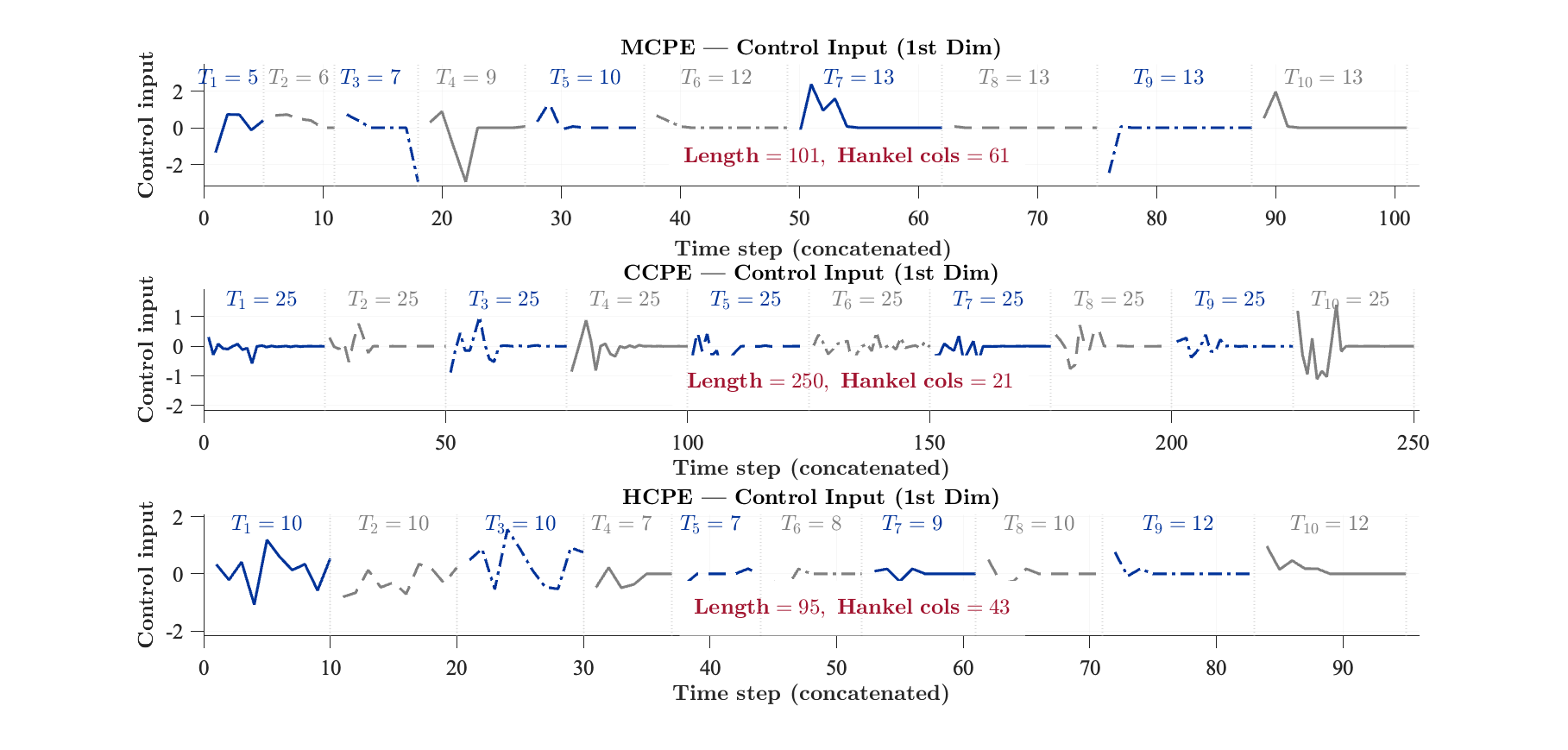}
	\caption {{Control input sequences generated under the MCPE, CCPE, and HCPE conditions, respectively, along with the total input lengths and the corresponding Hankel matrix column sizes.}}
	\label{control}
\end{figure*}

For the MCPE condition, the data sets can have varying dimensions but must satisfy \(\sum_{i=1}^p T_i \geq 50\) (see Theorem \ref{Proof of experimental design}). Control sequence lengths are randomly generated within the range [5, 14). For the CCPE condition, the ten data sets must have identical dimensions and satisfy \(T_0 \geq 14\) (see Theorem \ref{Proof of experimental design}), we set $T_0=25$. This larger choice is deliberate and provides a practical safety margin against measurement noise, finite-precision arithmetic, and unmodeled dynamics. It also increases the number of columns $ T_0 - L + 1$, which makes the ensuing least-squares and pseudoinverse computations more overdetermined and numerically stable. For the HCPE condition, partial dimensional consistency is allowed \(\bar{p}=3\), with the condition \(T_0 + \sum_{i=\bar{p}}^p T_i \geq 42\) (see Theorem \ref{Proof of experimental design}). Let \(T_0 = 10\), the remaining controller lengths are randomly generated within [5, 14). 

The input sequences for all three conditions are designed using the method in Section \ref{s4}. The initial state values are randomly generated within the range [-1, 1]. Additionally, the standard deviation of the process noise $\sigma_{w}$ is varied from $0$ to $0.1$. The weighting factors for each CPE condition are fixed as \(\alpha_i = 1\) for \(i = 1, 2, \dots, 10\). We solve for $\hat{G} $ in \eqref{G} using the \text{pinv} function from the standard MATLAB Linear Algebra Toolbox. The identification errors under various conditions are shown in Fig.  \ref{LS}(a)-(c). The results validate the effectiveness of the three synthesized conditions for least-squares identification. 

To highlight the significance of weighting factors, we examined scenarios where \(\alpha_1\) and \(\alpha_2-\alpha_{10}\) varied within the range (0, 30] under the MCPE condition. The results, corresponding to a noise level of \(\sigma_w = 0.05\), are presented in Fig. \ref{LS}(d). The red line represents the case where \(\alpha_1\) varies while \(\alpha_2-\alpha_{10}\) remain fixed at 1, whereas the blue line represents the case where \(\alpha_1\) is fixed at 1 and \(\alpha_2-\alpha_{10}\) vary synchronously. The findings indicate that significant disparities in signal magnitudes under the MCPE condition can degrade discrimination performance. Conversely, an appropriate selection of weighting factors effectively mitigates identification errors.

\subsection{State Feedback Controller Design Using Parametric Modeling Approach}
In this subsection, we apply the data-driven state feedback design method described in \cite{8933093} to the system defined by \eqref{batch reactor system} using the three CPE conditions. Specifically, taking the CCPE condition as an example. Utilizing the approach in \cite[Th. 3]{8933093}, the design challenge for stabilizing controllers is transformed into an LMI problem for any matirx $Q$:
\begin{align}\label{LMI}
	\begin{bmatrix}
		{{H_1^{cum}(\{{x_{i}^{[0,{T_0-1}]}\}_{i=1}^p})}Q}&{{H_1^{cum}(\{{x_{i}^{[1,{T_0}]}\}_{i=1}^p})}Q}\\
		{{{\left( {{H_1^{cum}(\{{x_{i}^{[1,{T_0}]}\}_{i=1}^p})}Q} \right)}^T}}&{{H_1^{cum}(\{{x_{i}^{[0,{T_0-1}]}\}_{i=1}^p})}Q}
	\end{bmatrix} > 0,
\end{align}
and the feedback control gain can be obtained by
\begin{align}\label{solve_lmi}
	K_{cum}=&{H_1^{cum}(\{{u_{i}^{[0,{T_0-1}]}\}_{i=1}^p})}Q \nonumber \\  
	&~~~~~~~~~~~\times {\left( {{H_1^{cum}(\{{x_{i}^{[0,{T_0-1}]}\}_{i=1}^p})}Q} \right)^{ - 1}}.
\end{align}

Similar to LS identification, \eqref{LMI} and \eqref{solve_lmi} are feasible if and only if the corresponding CPE conditions satisfy at least the $n+1$ order. For this example, we use the control signals generated in the previous subsection as a priori control inputs to generate corresponding input-state trajectories for the three CPE conditions of order $5$. Fig. \ref{control} presents the control input sequences generated under the MCPE, CCPE, and HCPE conditions, respectively. Each subplot corresponds to one excitation strategy and depicts the concatenated input signals of the first control component across ten trajectory segments. 

In the case of the CCPE condition, the collected input-state trajectories are synthesized into ${H_1^{cum}(\{{u_{i}^{[0,{T_0-1}]}\}_{i=1}^p})}$, ${H_1^{cum}(\{{x_{i}^{[0,{T_0-1}]}\}_{i=1}^p})}$, and ${H_1^{cum}(\{{x_{i}^{[1,{T_0}]}\}_{i=1}^p})}$. These matrices are then processed using MATLAB's LMI toolbox \cite{duan2013lmis} to compute the control gain $K_{cum}$ using \eqref{LMI} and \eqref{solve_lmi}:
$${K_{cum}} = \begin{bmatrix}
	{0.8480}&{ - 1.4693}&{0.1463}&{ - 1.5678}\\
	{4.0457}&{0.0394}&{3.2812}&{ - 1.4414}
\end{bmatrix}.$$

It can be verified that the closed-loop matrix $A+BK_{cum}$ is stable with a spectral radius of $0.5922$.

For the MCPE condition, the collected input-state trajectories are synthesized into ${H_1^{mos}(\{{u_{i}^{[0,{T_i-1}]}\}_{i=1}^p})}$, ${H_1^{mos}(\{{x_{i}^{[0,{T_i-1}]}\}_{i=1}^p})}$, and ${H_1^{mos}(\{{x_{i}^{[1,{T_i}]}\}_{i=1}^p})}$, and these matrices replace the corresponding matrices in  \eqref{LMI} and \eqref{solve_lmi}. Solving the resulting LMI yields the control gain $K_{mos}$:
$${K_{mos}} = \begin{bmatrix}
	{0.8218}&{ - 1.4683}&{0.1403}&{ - 1.5406}\\
	{3.9691}&{0.0432}&{3.2652}&{ - 1.3369}
\end{bmatrix}.$$

It can be verified that the closed-loop matrix $A+BK_{mos}$ is stable with a spectral radius of $0.5921$.

Finally for the HCPE condition, the resulting input-state trajectories are organized into ${H_1^{hyb}(\{{u_{i}^{[0,{T_i-1}]}\}_{i=1}^p})}$, ${H_1^{hyb}(\{{x_{i}^{[0,{T_i-1}]}\}_{i=1}^p})}$, and ${H_1^{hyb}(\{{x_{i}^{[1,{T_i}]}\}_{i=1}^p})}$,  replacing the corresponding matrices in \eqref{LMI} and \eqref{solve_lmi}. Solving yields a control gain $K_{hyb}$:
$${K_{hyb}} = \begin{bmatrix}
	{0.8416}&{ - 1.4690}&{0.1450}&{ - 1.5594}\\
	{4.0254}&{0.0399}&{3.2761}&{ - 1.4282}
\end{bmatrix}.$$

It can be verified that the closed-loop matrix $A+BK_{hyb}$ is stable with a spectral radius of $0.5942$.

\subsection{Data-Driven MPC Under Three CPE Conditions}
In this subsection, we apply the data-driven MPC method to a batch reactor system considered in the previous subsection.

It is feasible to directly employ Lemma \ref{lem.extend} for the MPC problem:
\begin{align*}
	\begin{array}{l}
		\mathop {\min }\limits_{g,{{\bar u}^{[1,T]}},{{\bar x}^{[1,T]}}} \sum\limits_{m = 1}^N {J\left( {{{\bar x}_m},{{\bar u}_m}} \right)}  \\ 
		s.t.\left\{ {\begin{array}{*{20}{c}}
				{\begin{bmatrix}
						{{{\bar u}^{[ - n + 1,N]}}(k)}\\
						{{{\bar x}^{[ - n + 1,N]}}(k)}
					\end{bmatrix} = \begin{bmatrix}
						{H_{N + n}^{cum}(\{u_i^{[0,{T_0} - 1]}\}_{i=1}^p)}\\
						{H_{N + n}^{cum}(\{x_i^{[0,{T_0} - 1]}\}_{i=1}^p)}
					\end{bmatrix}g(k)},\\
				{\begin{bmatrix}
						{{{\bar u}^{ [- n + 1,0]}}(k)}\\
						{{{\bar y}^{ [- n + 1,0]}}(k)}
					\end{bmatrix} = \begin{bmatrix}
						{{u^{[k - (n - 1),k]}}}\\
						{{x^{[k - (n - 1),k]}}}
				\end{bmatrix}},\\
				{{{\bar u}_m}(k) \in \mathbb{U},m \in \left\{ {1,2,...,N} \right\}},\\
				{{{\bar x}_m}(k) \in \mathbb{X},m \in \left\{ {1,2,...,N} \right\}},
		\end{array}} \right.
	\end{array}
\end{align*}
where $N$ is the predicted step size, $u^{[k - (n - 1),k]}$ and $x^{[k - (n - 1),k]}$ are the $k - (n - 1)$ moment to $k$ moment individual input-state trajectories, respectively, which are taken as the initial values at moment $k$. ${{\bar u}^{[1,N]}}$ and ${{{\bar x}^{[ 1,N]}}}$ are the trajectories to be optimized. The sets $\mathbb{U}$ and $\mathbb{X}$ describe feasible inputs and states, respectively. When the cost function of MPC is a quadratic stage cost, i.e., $${J\left( {{{\bar x}_m},{{\bar u}_m}} \right)} = ({{{\bar x}_{m}(k)}}-x^*)^TQ({{{\bar x}_{m}(k)}}-x^*)+{{{\bar u}_{m}^T(k)}}R{{{\bar u}_{m}(k)}},$$
where $Q$, $R>0$, the problem can be transformed into a standard quadratic programming (QP) problem. 

The primary goal is to track the set point of the system $x^* = [0\ 0\ 0 \ 0]^T$. We select a prediction horizon of $N = 5$ and set the weight matrices in the cost function as $Q = 3 \cdot I_n$ and $R =10^{-2}\cdot I_m$. There are no constraints imposed on the upper and lower bounds of the control values.

\begin{figure}[htbp]
	\centering
	\includegraphics[scale=0.4]{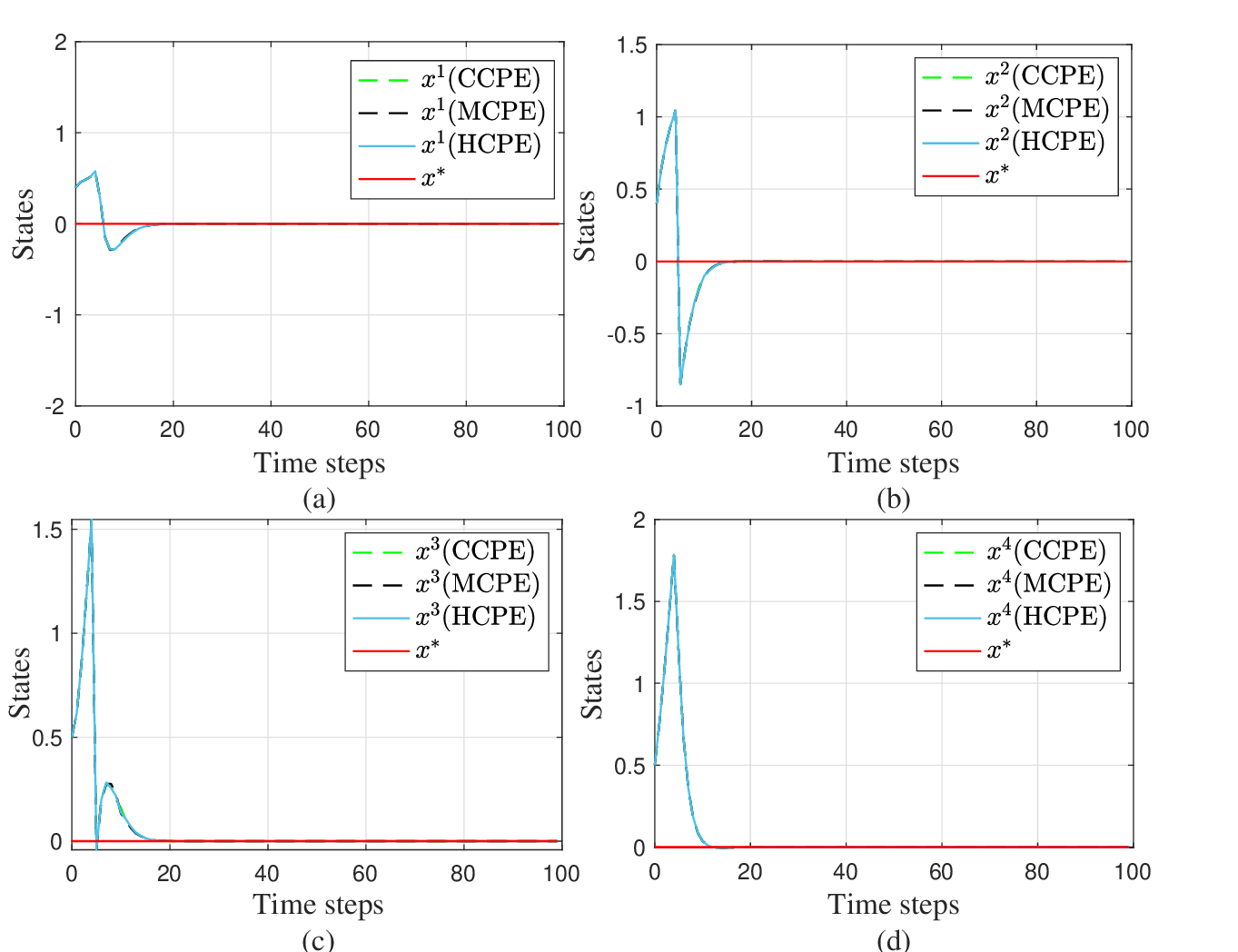}
	
	\caption{{Closed-loop states for data-driven MPC. (a). First state component. (b). Second state component. (c). Third state component. (d). Fourth state component.}}
	\label{mpc}
\end{figure}

We employed the proposed experimental design approach to collect the desired priori data satisfying three CPE conditions. For each condition, 10 sets of control sequences were collected, each with 30 steps. In Fig. \ref{mpc}, we present the closed-loop states obtained by implementing the data-driven MPC scheme under CCPE, MCPE, and HCPE conditions, respectively. Here, $x^i$, $i=1,2,...,n$ represents the $i$th state component of the closed-loop system. The outcomes demonstrate that the three methods exhibited nearly consistent (i.e., optimal) performance.

Furthermore, we compared the optimization times of the three methods, as shown in Table \ref{tab:running_times}. The MCPE method, which has the largest number of columns in the synthesis matrix and the highest computational complexity, required the longest computation time. Conversely, the CCPE method had the shortest computation time, while the HCPE method was intermediate. These results are consistent with our anticipated outcomes. 

\begin{table}[htbp]
	\centering
	\renewcommand{\arraystretch}{1.2}
	\setlength{\tabcolsep}{12pt}
	\caption{	\centering{Running Times of Different Conditions}}
	\label{tab:running_times}
	\begin{tabular}{lc}
		\toprule
		\textbf{Condition} & \textbf{Running Time (s)} \\
		\midrule
		CCPE & 0.1919 \\
		MCPE & 3.1657 \\
		HCPE & 1.8771 \\
		\bottomrule
	\end{tabular}
\end{table}

\subsection{Complexity Discussion Between MCPE and PE Conditions}\label{complexity discussion}

\textbf{\textit{Concrete numeric example (heterogeneous short trajectories):} }
Let $m=2$ and $L=5$. For a trajectory $u^{[0,T_0-1]}$, the Hankel matrix $H_L(u^{[0,T_0-1]})\in\mathbb{R}^{mL\times C_0}$ with column counts $C_0=T_0{-}L{+}1$. Recall that a single-trajectory PE condition requires
\begin{align*}
	T_0\ \ge\ (m+1)L-1 \;=\; 3\cdot 5-1 \;=\; 14.
\end{align*}
Suppose long experiments are infeasible, but we can repeatedly collect short trials.
Take five short trajectories with lengths
$$
T_1=7,\; T_2=7,\; T_3=6,\; T_4=6,\; T_5=5.
$$
Individually none satisfies $T_i\ge14$. The mosaic Hankel matrix  ${H_L^{mos}(\{u^{[0,T_i-1]}_i\}_{i=1}^{p})}\in\mathbb{R}^{mL\times C}$ with column counts $C=\sum_{i=1}^{N}\big(T_i-L+1\big)$. Their available column counts are
\begin{align*}
	T_1{-}L{+}1&=3,\; T_2{-}L{+}1=3,\; T_3{-}L{+}1=2, \\
	&T_4{-}L{+}1=2,\; T_5{-}L{+}1=1,
\end{align*}
so the mosaic Hankel matrix has
$$
C = 3+3+2+2+1 = 11 \;\;\ge\; mL=10.
$$
Under mild genericity (inputs not persistently collinear and weights $\alpha_i$ chosen to avoid near-cancellations), this suffices to achieve
$\rank(H^{mos}_L)=mL$.
Hence the MCPE test can be satisfied by one unified rank check on a $(mL)\times C = 10\times 11$ matrix, even though no single trajectory is long enough to satisfy PE.

\textbf{\textit{Complexity accounting:} } Let $\text{flops}_{(\cdot)}$ denote the leading-order cost of a rank test via singular value decomposition (SVD). 

For the PE condition, one rank test on $\mathbb{R}^{mL\times C_0}$ with $C_0=T_0{-}L{+}1\ge mL$. A typical cost model gives
$$
\text{flops}_{\text{PE}} \;\sim\; \mathcal{O}\big((mL)^2 C_0\big).
$$
However, when a long trajectory is not available upfront, one often repeats experiments and re-checks rank $K$ times until Definition \ref{Persistently Exciting Condition} is met, giving a cumulative
\begin{align}\label{PErepeat}
	\text{flops}_{\text{PE, repeated}} \;\sim\; \sum_{k=1}^{K}\mathcal{O}\big((mL)^2 C_k\big),
\end{align}
with added experimental overhead for each trial.

For the MCPE condition, a single rank test on $\mathbb{R}^{mL\times C}$ with $C=\sum_i (T_i{-}L{+}1)_+$:
\begin{align}\label{MCPEcost}
	\text{flops}_{\text{MCPE}} \;\sim\; \mathcal{O}\big((mL)^2 C\big).
\end{align}

Comparing \eqref{PErepeat} and \eqref{MCPEcost}, repeated PE testing becomes more expensive once
\begin{align}\label{threshold}
	\sum_{k=1}^K C_k > C,
\end{align}
that is, when the cumulative number of columns generated by repeated trials exceeds that of the single mosaic matrix. If each trial produces of PE condition on average $\bar c$ columns, the threshold occurs at
\begin{equation}\label{Kthreshold}
	K_{\mathrm{th}} = \Big\lceil \tfrac{C}{\bar c} \Big\rceil.
\end{equation}
In the previous example, if each failed PE attempt produces is $\bar c=C_0=10$ columns, then by \eqref{Kthreshold} the threshold is $K_{\mathrm{th}}=\lceil 11/10\rceil=2$. Hence after two repeated PE trials the cumulative cost exceeds that of a single MCPE check.

This illustrates that, although MCPE involves a wider matrix than a one-shot PE check, it avoids repeated failed trials and preprocessing, thereby reducing the overall computational and experimental burden.

\section{Conclution}\label{s7}
In this paper, we have developed and analyzed three collectively persistently exciting (CPE) conditions, each incorporating weight factors, from the perspectives of same-dimensional, different-dimensional, and partially same-dimensional data. We argue that these three CPE conditions offer valuable insights for designing data-driven methodologies tailored to diverse multi-signal control scenarios, including those involving signals with insufficient informativeness. We have explored the interrelations between these conditions and examined their rank properties within the framework of linear time-invariant systems.  All three conditions successfully extend Willems’ fundamental lemma. To address the challenges of insufficient information in signal segments, we have proposed open-loop experimental design methods tailored to each CPE condition, enabling the synthesis of the required excitation conditions both offline or online. Illustrative examples show that these conditions lead to satisfactory results in both model identification and the development of data-driven controllers. 

Several topics for future research remain open. An interesting but challenging extension lies in devising data-driven control methods in a distributed framework, as opposed to a centralized one, while ensuring adherence to the three CPE conditions with multiple control signals. Furthermore, we note that weighting factors have an impact on the performance and effectiveness of data-driven methods. Therefore, how to systematically design optimal weighting factors for data-driven problems is another direction worth exploring. In addition, it is of interest to investigate how different experiment design methods influence parameter uncertainties and control performance, thereby integrating uncertainty-aware objectives into the proposed framework.

\section*{Appendix}
\subsection*{A. Proof of Theorem 1}
1). If $\{z_{i}^{[0,{T_i-1}]}\}_{i=1}^p$ are CCPE, then $T_0=T_1=T_2=...=T_{p}$ and ${H_L^{cum}(\{{z_{i}^{[0,{T_i-1}]}\}_{i=1}^p})}{{H_L^{cum}(\{{z_{i}^{[0,{T_i-1}]}\}_{i=1}^p})}}^\top>0$, i.e., 
\begin{align}\label{inequality1}
	&\sum\limits_{i = 1}^p {\left( {{\alpha _i}{H_L}({z_{i}^{[0,{T_i} - 1]}})\sum\limits_{i \ne j = 1}^p {{\alpha _j}H_L^T({z_{j}^{[0,{T_j} - 1]}})} } \right)}\nonumber \\
	&~~~~~~~~~~+\underbrace {\alpha _i^2\sum\limits_{i = 1}^p {{H_L}({z_{i}^{[0,{T_i} - 1]}})H_L^T({z_{i}^{[0,{T_i} - 1]}})} }_{M_1} >0. &
\end{align}

In order to establish the full row rank of ${H_L^{mos}(\{{z_{i}^{[0,{T_i-1}]}\}_{i=1}^p})}$, it is sufficient to show the positivity of $M_1$. Given $\sum\limits_{i = 1}^p {{H_L}({z_{i}^{[0,{T_i} - 1]}})H_L^T({z_{i}^{[0,{T_i} - 1]}})}\geq0$ and $\alpha _i \neq 0$, if there exists a non-zero vector $x$ such that $x\in\ker M_1$, then we must have $H_L^T({z_{1}^{[0,{T_i} - 1]}})x =  \cdots H_L^T({z_{p}^{[0,{T_i} - 1]}})x = 0$, implying $$x\in\ker\sum\limits_{i = 1}^p {\left( {{\alpha _i}{H_L}({z_{i}^{[0,{T_i} - 1]}})\sum\limits_{i \ne j = 1}^p {{\alpha _j}H_L^T({z_{j}^{[0,{T_j} - 1]}})} } \right)}.$$ 

This conclusion contradicts inequality \eqref{inequality1}. Therefore $M_1>0$. Consequently, $\{z_{i}^{[0,{T_i-1}]}\}_{i=1}^p$ are also MCPE.

Since ${H_L^{cum}(\{{z_{i}^{[0,{T_i-1}]}\}_{i=1}^p})} = {H_L^{cum}(\{{z_{i}^{[0,{T_i-1}]}\}_{i=1}^{\bar{p}}})} + {H_L^{cum}(\{{z_{i}^{[0,{T_i-1}]}\}_{i=\bar{p}+1}^{p}})} $, the method employed above to establish the full row rank of ${H_L^{mos}(\{{z_{i}^{[0,{T_i-1}]}\}_{i=1}^p})}$ from the full row rank matrix ${H_L^{cum}(\{{z_{i}^{[0,{T_i-1}]}\}_{i=1}^p})}$ is also applicable to determining the full row rank of ${\begin{bmatrix}
		{H_L^{cum}(\{z^{[0,T_0-1]}_i\}_{i=1}^{\bar{p}})} &{H_L^{mos}(\{z^{[0,T_i-1]}_i\}_{i=\bar{p}+1}^{p})}
\end{bmatrix}} $ from the full row rank matrix ${H_L^{cum}(\{{z_{i}^{[0,{T_i-1}]}\}_{i=1}^{\bar{p}}})} + {H_L^{cum}(\{{z_{i}^{[0,{T_i-1}]}\}_{i=\bar{p}+1}^{p}})} $. Therefore $\{z_{i}^{[0,{T_i-1}]}\}_{i=1}^p$ are also HCPE.

2). \romannumeral1).  If $\{z_{i}^{[0,{T_i-1}]}\}_{i=1}^p$ are MCPE and $T_0=T_1=T_2=...=T_{p}$, we have
${H_L^{cum}(\{{z_{i}^{[0,{T_i-1}]}\}_{i=1}^p})} ={H_L^{mos}(\{{z_{i}^{[0,{T_i-1}]}\}_{i=1}^p})} ({\textbf{1}_p} \otimes {I_{{T_0} - L + 1}})$. Then, 
\begin{align*}
	&\text{rank}\ {H_L^{cum}(\{{z_{i}^{[0,{T_i-1}]}\}_{i=1}^p})} = \text{rank}\ {H_L^{mos}(\{{z_{i}^{[0,{T_i-1}]}\}_{i=1}^p})} \nonumber\\
	&~~~-~ \text{dim} [\text{im} \, {H_L^{mos}(\{{z_{i}^{[0,{T_i-1}]}\}_{i=1}^p})^\top} \cap \text{im} ({{\bf{1}}_p} \otimes {I_{{T_0} - L + 1}})^{\perp}].&
\end{align*} 
Thus, we have $$\text{rank}\ {H_L^{cum}(\{{z_{i}^{[0,{T_i-1}]}\}_{i=1}^p})} = \text{rank}\ {H_L^{mos}(\{{z_{i}^{[0,{T_i-1}]}\}_{i=1}^p})} $$ if $\text{im} \, {H_L^{mos}(\{{z_{i}^{[0,{T_i-1}]}\}_{i=1}^p})^\top} 
\cap \text{leftker}\, {\textbf{1}_p} \otimes {I_{{T_0} - L + 1}}=\{0\}$, i.e., $\{z_{i}^{[0,{T_i-1}]}\}_{i=1}^p$ are also CCPE.

2). \romannumeral2). If $T_0=T_1=T_2=...=T_{\bar{p}}$, we have 
\begin{align*}
	&{H_L^{hyb}(\{{z_{i}^{[0,{T_i-1}]}\}_{i=1}^p})} \nonumber \\
	&~~~~= {H_L^{mos}(\{{z_{i}^{[0,{T_i-1}]}\}_{i=1}^{\bar{p}}})}(\begin{bmatrix}
		{{\textbf{1}_{\bar p}}}&{{0_{\bar p \times (p - \bar p)}}}\\
		{{0_{p - \bar p}}}&{{I_{p - \bar p}}}
	\end{bmatrix} \otimes {I_{{T_0} - L + 1}}).
\end{align*}

Similarly to the above, if additionally, $\text{im}\,  {{H_L^{mos}(\{{z_{i}^{[0,{T_i-1}]}\}_{i=1}^{\bar{p}}})}^\top}
\cap \text{leftker}\, {\begin{bmatrix}
		{{\textbf{1}_{\bar p}}}&{{0_{\bar p \times (p - \bar p)}}}\\
		{{0_{p - \bar p}}}&{{I_{p - \bar p}}}
\end{bmatrix}} \otimes {I_{{T_0} - L + 1}}=\{0\}$, then $\{z_{i}^{[0,{T_i-1}]}\}_{i=1}^p$ are HCPE. 

3). \romannumeral1). If $\{z_{i}^{[0,{T_i-1}]}\}_{i=1}^p$ are HCPE, then $T_0=T_1=...=T_{\bar{p}}$ and ${H_L^{hyb}(\{z^{[0,T_i-1]}_i\}_{i=1}^{p})}{H_L^{hyb}(\{z^{[0,T_i-1]}_i\}_{i=1}^{p})}^\top>0$, i.e., ${M_1} + \sum\limits_{i = 1}^{\bar p} {\left( {{\alpha _i}{H_L}({z_{i}^{[0,{T_0} - 1]}})\sum\limits_{i \ne j = 1}^{\bar p} {{\alpha _j}H_L^\top({z_{j}^{[0,{T_0} - 1]}})} } \right)}>0$. Similar to the proof of conclusion 1), we can deduce  $M_1 > 0$, and thus $\{z_{i}^{[0,{T_i-1}]}\}_{i=1}^p$ are also MCPE. 

3). \romannumeral2). If $T_0=T_{\bar{p}+1}=...=T_{p}$, then ${H_L^{hyb}(\{{z_{i}^{[0,{T_i-1}]}\}_{i=1}^p})}$ can be regarded as a special case of ${H_L^{mos}(\{{z_{i}^{[0,{T_i-1}]}\}_{i=1}^p})} $ in conclusion 2). \romannumeral1). The proof for this case is evidently straightforward.  \hfill $\blacksquare$ 

\subsection*{B. Proof of Lemma 3.1}

We begin by demonstrating that  $$\text{rank}\, \begin{bmatrix}
	{H_1^{cum}(\{{x_{i}^{[0,{T_i-L}]}\}_{i=1}^p})}\\
	{H_L^{cum}(\{{u_{i}^{[0,{T_i-1}]}\}_{i=1}^p})}
\end{bmatrix} = n + mL,$$ i.e., $\begin{bmatrix}
	{H_1^{cum}(\{{x_{i}^{[0,{T_i-L}]}\}_{i=1}^p})}\\
	{H_L^{cum}(\{{u_{i}^{[0,{T_i-1}]}\}_{i=1}^p})}
\end{bmatrix}$ achieves full row rank when $\{u_{i}^{[0,T_i-1]}\}_{i=1}^p$ is CCPE. 

Let $[r_x \  r_u] \in \text{leftker}\left[ {\begin{array}{*{20}{c}}
		{H_1^{cum}(\{{x_{i}^{[0,{T_i-L}]}\}_{i=1}^p})}\\
		{H_L^{cum}(\{{u_{i}^{[0,{T_i-1}]}\}_{i=1}^p})}
\end{array}} \right]$, where $r_x^\top \in \mathbb{R}^n$, $r_u^\top \in \mathbb{R}^{mL}$. In the context of system \eqref{Sysform. (1)}, we konw that
\begin{align*}
	&\begin{bmatrix}
		{H_1^{cum}(\{{x_{i}^{[1,{T_i-L-n+1}]}\}_{i=1}^p})}\nonumber\\
		{H_L^{cum}(\{{u_{i}^{[1,{T_i-n}]}\}_{i=1}^p})}
	\end{bmatrix}=\\
	&  \begin{bmatrix}
		{\begin{bmatrix}
				A&B
		\end{bmatrix}}&{{0_{n \times m(n + L - 1)}}}\\
		{{0_{mL \times (n + m)}}}&{\begin{bmatrix}
				I_{mL}&{{0}}
		\end{bmatrix}}
	\end{bmatrix}\begin{bmatrix}
		{H_1^{cum}(\{{x_{i}^{[0,{T_i-L-n}]}\}_{i=1}^p})}\\
		{H_{L+n}^{cum}(\{{u_{i}^{[0,{T_i-1}]}\}_{i=1}^p})}
	\end{bmatrix}
\end{align*}
\centerline{$\vdots$}
\begin{align*}
	&\begin{bmatrix}
		{H_1^{cum}(\{{x_{i}^{[n,{T_i-L}]}\}_{i=1}^p})}\\
		{H_L^{cum}(\{{u_{i}^{[n,{T_i-1}]}\}_{i=1}^p})}
	\end{bmatrix}= \\
	&\begin{bmatrix}
		{\begin{bmatrix}
				{{A^n}}&{{A^{n - 1}}B}& \cdots &B
		\end{bmatrix}}&{{0}}\\
		{{0_{mL \times n(1 + m)}}}&I_{mL}
	\end{bmatrix}\begin{bmatrix}
		{H_1^{cum}(\{{x_{i}^{[0,{T_i-L-n}]}\}_{i=1}^p})}\\
		{H_{L+n}^{cum}(\{{u_{i}^{[0,{T_i-1}]}\}_{i=1}^p})}
	\end{bmatrix}.
\end{align*}

Since $[r_x \ r_u]$ also resides in the left kernel of the matrices on the left side of the above equations, a simultaneous left-multiplication of all equations by $[r_x \ r_u]$ yields
\begin{align*}
	R_x^u\begin{bmatrix}
		{H_1^{cum}(\{{x_{i}^{[0,{T_i-L-n}]}\}_{i=1}^p})}\\
		{H_{L+n}^{cum}(\{{u_{i}^{[0,{T_i-1}]}\}_{i=1}^p})}
	\end{bmatrix}= {0_{(n + 1) \times ({T_i} - L - n + 1)}},
\end{align*}
where $R_x^u=\begin{bmatrix}
	{{r_x}}&{{r_u}}&{}&{}&{}\\
	{{r_x}A}&{{r_x}B}&{{r_u}}&{}&{}\\
	\vdots & \ddots & \ddots & \ddots &{}\\
	{{r_x}{A^n}}&{{r_x}{A^{n - 1}}B}& \cdots &{{r_x}B}&{{r_u}}
\end{bmatrix}$. 

Through CCPE condition, one obtains $\text{rank}\,	{H_{L+n}^{cum}(\{{u_{i}^{[0,{T_i-1}]}\}_{i=1}^p})}=m(n+L)$. There must have $\text{dim}(\text{leftker}\begin{bmatrix}
	{H_1^{cum}(\{{x_{i}^{[0,{T_i-L-n}]}\}_{i=1}^p})}\\
	{H_{L+n}^{cum}(\{{u_{i}^{[0,{T_i-1}]}\}_{i=1}^p})}
\end{bmatrix})\leq n$, thus $\text{rank}(R_x^u)\leq n$. Since $R_x^u$ is a lower triangular block matrix with $n + 1$ rows, it follows that at least one of these rows causes $R_x^u$ to drop rank. Consequently, $r_u$ and $r_xB$ must both be $0$ vectors. At this point, there are two cases for letting $R_x^u$ unrank to $n$.

1. There exists a nonzero constant $\tau_0$ such that $\tau_0 r_x=r_xA$. Then we have $r_xAB=\tau_0 r_xB=0$, and further recursion yields $r_xA^{n-1}B=r_xA^{n-2}B\cdots= r_xB=0$. The reachability of $(A,B)$ ensures that $r_x=0$.

2. letting $r_xA^jB=0$, $j=1,2,..,n-1$ and there exists a nonzero constant $\tau_j$ such that $\tau_jr_xA^j=r_xA^{j+1}$, in which case the same result is obtained with $r_xA^{n-1}B=r_xA^{n-2}B\cdots= r_xB=0$, and thus $r_x=0$. 

In summary, $\text{rank}(R_x^u)\leq n$ implies that $r_x=r_u=0$, and thus the matirx $\begin{bmatrix}
	{H_1^{cum}(\{{x_{i}^{[0,{T_i-L}]}\}_{i=1}^p})}\\
	{H_L^{cum}(\{{u_{i}^{[0,{T_i-1}]}\}_{i=1}^p})}
\end{bmatrix}$ has full row rank.

Similarly, employing a comparable approach, we can establish that $\begin{bmatrix}
	{H_1^{mos}(\{{x_{i}^{[0,{T_i-L}]}\}_{i=1}^p})}\\
	{H_L^{mos}(\{{u_{i}^{[0,{T_i-1}]}\}_{i=1}^p})}
\end{bmatrix}$ or $\begin{bmatrix}
	{H_1^{hyb}(\{{x_{i}^{[0,{T_i-L}]}\}_{i=1}^p})}\\
	{H_L^{hyb}(\{{u_{i}^{[0,{T_i-1}]}\}_{i=1}^p})}
\end{bmatrix}$ achieves full row rank when $\{u_{i}^{[0,{T_i-1}]}\}_{i=1}^p$ is MCPE or HCPE of order $L+n$. We will not reiterate this here. \hfill $\blacksquare$ 
\bibliography{ref.bib} 

\begin{thebibliography}{41}
\expandafter\ifx\csname natexlab\endcsname\relax\def\natexlab#1{#1}\fi
\providecommand{\url}[1]{\texttt{#1}}
\providecommand{\href}[2]{#2}
\providecommand{\path}[1]{#1}
\providecommand{\DOIprefix}{doi:}
\providecommand{\ArXivprefix}{arXiv:}
\providecommand{\URLprefix}{URL: }
\providecommand{\Pubmedprefix}{pmid:}
\providecommand{\doi}[1]{\href{http://dx.doi.org/#1}{\path{#1}}}
\providecommand{\Pubmed}[1]{\href{pmid:#1}{\path{#1}}}
\providecommand{\bibinfo}[2]{#2}
\ifx\xfnm\relax \def\xfnm[#1]{\unskip,\space#1}\fi
\bibitem[{Borwein and Lewis(2006)}]{L_convex}
\bibinfo{author}{Borwein, J.}, \bibinfo{author}{Lewis, A.},
  \bibinfo{year}{2006}.
\newblock \bibinfo{title}{Convex Analysis}.
\newblock \bibinfo{publisher}{Springer New York}.
\bibitem[{Chu et~al.(2016)Chu, Gao and Zhang}]{2016output}
\bibinfo{author}{Chu, H.}, \bibinfo{author}{Gao, L.}, \bibinfo{author}{Zhang,
  W.}, \bibinfo{year}{2016}.
\newblock \bibinfo{title}{Distributed adaptive containment control of
  heterogeneous linear multi-agent systems: an output regulation approach}.
\newblock \bibinfo{journal}{IET Control Theory $\&$ Applications}
  \bibinfo{volume}{10}, \bibinfo{pages}{95--102}.
\bibitem[{El~Ghor and Aggoune(2020)}]{energy-1}
\bibinfo{author}{El~Ghor, H.}, \bibinfo{author}{Aggoune, E.H.M.},
  \bibinfo{year}{2020}.
\newblock \bibinfo{title}{Energy efficient scheduler of aperiodic jobs for
  real-time embedded systems}.
\newblock \bibinfo{journal}{International Journal of Automation and Computing}
  \bibinfo{volume}{17}, \bibinfo{pages}{733--743}.
\bibitem[{Ge et~al.(2021)Ge, Han, Zhang and Ding}]{2021dynamic-survey}
\bibinfo{author}{Ge, X.}, \bibinfo{author}{Han, Q.}, \bibinfo{author}{Zhang,
  X.}, \bibinfo{author}{Ding, D.}, \bibinfo{year}{2021}.
\newblock \bibinfo{title}{Dynamic event-triggered control and estimation: A
  survey}.
\newblock \bibinfo{journal}{International Journal of Automation and Computing}
  \bibinfo{volume}{18}, \bibinfo{pages}{857--886}.
\bibitem[{Haghshenas et~al.(2015)Haghshenas, Badamchizadeh and
  Baradarannia}]{2015containment}
\bibinfo{author}{Haghshenas, H.}, \bibinfo{author}{Badamchizadeh, M.A.},
  \bibinfo{author}{Baradarannia, M.}, \bibinfo{year}{2015}.
\newblock \bibinfo{title}{Containment control of heterogeneous linear
  multi-agent systems}.
\newblock \bibinfo{journal}{Automatica} \bibinfo{volume}{54},
  \bibinfo{pages}{210--216}.
\bibitem[{Horn and Johnson(2012)}]{L_Disc-Theorem}
\bibinfo{author}{Horn, R.A.}, \bibinfo{author}{Johnson, C.R.},
  \bibinfo{year}{2012}.
\newblock \bibinfo{title}{Matrix analysis}.
\newblock \bibinfo{publisher}{Cambridge university press}.
\bibitem[{Hu et~al.(2016)Hu, Liu and Feng}]{ET-U-3}
\bibinfo{author}{Hu, W.}, \bibinfo{author}{Liu, L.}, \bibinfo{author}{Feng,
  G.}, \bibinfo{year}{2016}.
\newblock \bibinfo{title}{Output consensus of heterogeneous linear multi-agent
  systems by distributed event-triggered/self-triggered strategy}.
\newblock \bibinfo{journal}{IEEE Transactions on Cybernetics}
  \bibinfo{volume}{47}, \bibinfo{pages}{1914--1924}.
\bibitem[{Isidori and Byrnes(1990)}]{2023_2}
\bibinfo{author}{Isidori, A.}, \bibinfo{author}{Byrnes, C.},
  \bibinfo{year}{1990}.
\newblock \bibinfo{title}{Output regulation of nonlinear systems}.
\newblock \bibinfo{journal}{IEEE Transactions on Automatic Control}
  \bibinfo{volume}{35}, \bibinfo{pages}{131--140}.
\bibitem[{Jadbabaie et~al.(2003)Jadbabaie, Lin and Morse}]{DY6_2003}
\bibinfo{author}{Jadbabaie, A.}, \bibinfo{author}{Lin, J.},
  \bibinfo{author}{Morse, A.S.}, \bibinfo{year}{2003}.
\newblock \bibinfo{title}{Coordination of groups of mobile autonomous agents
  using nearest neighbor rules}.
\newblock \bibinfo{journal}{IEEE Transactions on Automatic Control}
  \bibinfo{volume}{48}, \bibinfo{pages}{988--1001}.
\bibitem[{Kamalapurkar et~al.(2015)Kamalapurkar, Dinh, Bhasin and
  Dixon}]{2023_1}
\bibinfo{author}{Kamalapurkar, R.}, \bibinfo{author}{Dinh, H.},
  \bibinfo{author}{Bhasin, S.}, \bibinfo{author}{Dixon, W.E.},
  \bibinfo{year}{2015}.
\newblock \bibinfo{title}{Approximate optimal trajectory tracking for
  continuous-time nonlinear systems}.
\newblock \bibinfo{journal}{Automatica} \bibinfo{volume}{51},
  \bibinfo{pages}{40--48}.
\bibitem[{Li et~al.(2021)Li, Sun, Tang and Karimi}]{ET-U-1}
\bibinfo{author}{Li, X.}, \bibinfo{author}{Sun, Z.}, \bibinfo{author}{Tang,
  Y.}, \bibinfo{author}{Karimi, H.R.}, \bibinfo{year}{2021}.
\newblock \bibinfo{title}{Adaptive event-triggered consensus of multiagent
  systems on directed graphs}.
\newblock \bibinfo{journal}{IEEE Transactions on Automatic Control}
  \bibinfo{volume}{66}, \bibinfo{pages}{1670--1685}.
\bibitem[{Li et~al.(2022)Li, Li and Tong}]{20230403add}
\bibinfo{author}{Li, Y.}, \bibinfo{author}{Li, Y.X.}, \bibinfo{author}{Tong,
  S.}, \bibinfo{year}{2022}.
\newblock \bibinfo{title}{Event-based finite-time control for nonlinear
  multi-agent systems with asymptotic tracking}.
\newblock \bibinfo{journal}{IEEE Transactions on Automatic Control} ,
  \bibinfo{pages}{http://dx.doi.org/10.1109/TAC.2022.3197562. Early Access}.
\bibitem[{Liu et~al.(2012)Liu, Xie and Wang}]{2012necessary}
\bibinfo{author}{Liu, H.}, \bibinfo{author}{Xie, G.}, \bibinfo{author}{Wang,
  L.}, \bibinfo{year}{2012}.
\newblock \bibinfo{title}{Necessary and sufficient conditions for containment
  control of networked multi-agent systems}.
\newblock \bibinfo{journal}{Automatica} \bibinfo{volume}{48},
  \bibinfo{pages}{1415--1422}.
\bibitem[{L{\"u} et~al.(2020)L{\"u}, He, Han, Ge and Peng}]{nonlinear_L2020}
\bibinfo{author}{L{\"u}, H.}, \bibinfo{author}{He, W.}, \bibinfo{author}{Han,
  Q.L.}, \bibinfo{author}{Ge, X.}, \bibinfo{author}{Peng, C.},
  \bibinfo{year}{2020}.
\newblock \bibinfo{title}{Finite-time containment control for nonlinear
  multi-agent systems with external disturbances}.
\newblock \bibinfo{journal}{Information Sciences} \bibinfo{volume}{512},
  \bibinfo{pages}{338--351}.
\bibitem[{Lui et~al.(2021)Lui, Petrillo and Santini}]{6bu-6}
\bibinfo{author}{Lui, D.G.}, \bibinfo{author}{Petrillo, A.},
  \bibinfo{author}{Santini, S.}, \bibinfo{year}{2021}.
\newblock \bibinfo{title}{Distributed model reference adaptive containment
  control of heterogeneous multi-agent systems with unknown uncertainties and
  directed topologies}.
\newblock \bibinfo{journal}{Journal of the Franklin Institute}
  \bibinfo{volume}{358}, \bibinfo{pages}{737--756}.
\bibitem[{Miskowicz(2018)}]{energy-2}
\bibinfo{author}{Miskowicz, M.}, \bibinfo{year}{2018}.
\newblock \bibinfo{title}{Event-based control and signal processing}.
\newblock \bibinfo{publisher}{CRC press}.
\bibitem[{Nowzari et~al.(2019)Nowzari, Garcia and
  Cort{\'e}s}]{2019event-survey}
\bibinfo{author}{Nowzari, C.}, \bibinfo{author}{Garcia, E.},
  \bibinfo{author}{Cort{\'e}s, J.}, \bibinfo{year}{2019}.
\newblock \bibinfo{title}{Event-triggered communication and control of
  networked systems for multi-agent consensus}.
\newblock \bibinfo{journal}{Automatica} \bibinfo{volume}{105},
  \bibinfo{pages}{1--27}.
\bibitem[{Plemmons(1977)}]{1977M-matrix}
\bibinfo{author}{Plemmons, R.J.}, \bibinfo{year}{1977}.
\newblock \bibinfo{title}{{M-matrix characterizations. I—nonsingular
  M-matrices}}.
\newblock \bibinfo{journal}{Linear Algebra and its Applications}
  \bibinfo{volume}{18}, \bibinfo{pages}{175--188}.
\bibitem[{Qian and Wan(2021)}]{DY3-2021-2}
\bibinfo{author}{Qian, Y.}, \bibinfo{author}{Wan, Y.}, \bibinfo{year}{2021}.
\newblock \bibinfo{title}{Design of distributed adaptive event-triggered
  consensus control strategies with positive minimum inter-event times}.
\newblock \bibinfo{journal}{Automatica} \bibinfo{volume}{133},
  \bibinfo{pages}{no.109837}.
\bibitem[{Qin et~al.(2017)Qin, Ma, Shi and Wang}]{DY6_2016consensus}
\bibinfo{author}{Qin, J.}, \bibinfo{author}{Ma, Q.}, \bibinfo{author}{Shi, Y.},
  \bibinfo{author}{Wang, L.}, \bibinfo{year}{2017}.
\newblock \bibinfo{title}{Recent advances in consensus of multi-agent systems:
  A brief survey}.
\newblock \bibinfo{journal}{IEEE Transactions on Industrial Electronics}
  \bibinfo{volume}{64}, \bibinfo{pages}{4972--4983}.
\bibitem[{Rami et~al.(2001)Rami, Chen, Moore and Zhou}]{Riccati2}
\bibinfo{author}{Rami, M.A.}, \bibinfo{author}{Chen, X.},
  \bibinfo{author}{Moore, J.B.}, \bibinfo{author}{Zhou, X.Y.},
  \bibinfo{year}{2001}.
\newblock \bibinfo{title}{Solvability and asymptotic behavior of generalized
  riccati equations arising in indefinite stochastic lq controls}.
\newblock \bibinfo{journal}{IEEE Transactions on Automatic Control}
  \bibinfo{volume}{46}, \bibinfo{pages}{428--440}.
\bibitem[{Rusnak(1988)}]{Riccati1}
\bibinfo{author}{Rusnak, I.}, \bibinfo{year}{1988}.
\newblock \bibinfo{title}{Almost analytic representation for the solution of
  the differential matrix riccati equation}.
\newblock \bibinfo{journal}{IEEE Transactions on Automatic Control}
  \bibinfo{volume}{33}, \bibinfo{pages}{191--193}.
\bibitem[{Sun et~al.(2021)Sun, Zou, Guo and Xiang}]{DY1_2021event_S1}
\bibinfo{author}{Sun, Y.}, \bibinfo{author}{Zou, W.}, \bibinfo{author}{Guo,
  J.}, \bibinfo{author}{Xiang, Z.}, \bibinfo{year}{2021}.
\newblock \bibinfo{title}{Containment control for heterogeneous nonlinear
  multi-agent systems under distributed event-triggered schemes}.
\newblock \bibinfo{journal}{Frontiers of Information Technology $\&$ Electronic
  Engineering} \bibinfo{volume}{22}, \bibinfo{pages}{107--119}.
\bibitem[{Tang et~al.(2018)Tang, Deng and Hong}]{modular_Tang2018}
\bibinfo{author}{Tang, Y.}, \bibinfo{author}{Deng, Z.}, \bibinfo{author}{Hong,
  Y.}, \bibinfo{year}{2018}.
\newblock \bibinfo{title}{Optimal output consensus of high-order multiagent
  systems with embedded technique}.
\newblock \bibinfo{journal}{IEEE Transactions on Cybernetics}
  \bibinfo{volume}{49}, \bibinfo{pages}{1768--1779}.
\bibitem[{Wang et~al.(2021a)Wang, Wang, Wang and Wang}]{DY5_2020_NotFully}
\bibinfo{author}{Wang, D.}, \bibinfo{author}{Wang, Z.}, \bibinfo{author}{Wang,
  Z.}, \bibinfo{author}{Wang, W.}, \bibinfo{year}{2021}a.
\newblock \bibinfo{title}{Design of hybrid event-triggered containment
  controllers for homogeneous and heterogeneous multiagent systems}.
\newblock \bibinfo{journal}{IEEE Transactions on Cybernetics}
  \bibinfo{volume}{51}, \bibinfo{pages}{4885--4896}.
\bibitem[{Wang et~al.(2020a)Wang, Ni, Liu and Chen}]{6bu-5}
\bibinfo{author}{Wang, F.}, \bibinfo{author}{Ni, Y.}, \bibinfo{author}{Liu,
  Z.}, \bibinfo{author}{Chen, Z.}, \bibinfo{year}{2020}a.
\newblock \bibinfo{title}{Containment control for general second-order
  multiagent systems with switched dynamics}.
\newblock \bibinfo{journal}{IEEE Transactions on Cybernetics}
  \bibinfo{volume}{50}, \bibinfo{pages}{550--560}.
\bibitem[{Wang et~al.(2020b)Wang, Ni, Liu and Chen}]{6bu-4}
\bibinfo{author}{Wang, F.}, \bibinfo{author}{Ni, Y.}, \bibinfo{author}{Liu,
  Z.}, \bibinfo{author}{Chen, Z.}, \bibinfo{year}{2020}b.
\newblock \bibinfo{title}{Fully distributed containment control for
  second-order multi-agent systems with communication delay}.
\newblock \bibinfo{journal}{ISA Transactions} \bibinfo{volume}{99},
  \bibinfo{pages}{123--129}.
\bibitem[{Wang et~al.(2021b)Wang, Ren, Yu and Zhang}]{DY6_2021consensus}
\bibinfo{author}{Wang, H.}, \bibinfo{author}{Ren, W.}, \bibinfo{author}{Yu,
  W.}, \bibinfo{author}{Zhang, D.}, \bibinfo{year}{2021}b.
\newblock \bibinfo{title}{Fully distributed consensus control for a class of
  disturbed second-order multi-agent systems with directed networks}.
\newblock \bibinfo{journal}{Automatica} \bibinfo{volume}{132},
  \bibinfo{pages}{no.109816}.
\bibitem[{Wang et~al.(2017)Wang, Fu and Wang}]{nonlinear_Wang2017}
\bibinfo{author}{Wang, Q.}, \bibinfo{author}{Fu, J.}, \bibinfo{author}{Wang,
  J.}, \bibinfo{year}{2017}.
\newblock \bibinfo{title}{Fully distributed containment control of high-order
  multi-agent systems with nonlinear dynamics}.
\newblock \bibinfo{journal}{Systems \& Control Letters} \bibinfo{volume}{99},
  \bibinfo{pages}{33--39}.
\bibitem[{Wang et~al.(2022)Wang, Liu, Wu and Li}]{2021modular}
\bibinfo{author}{Wang, X.}, \bibinfo{author}{Liu, W.}, \bibinfo{author}{Wu,
  Q.}, \bibinfo{author}{Li, S.}, \bibinfo{year}{2022}.
\newblock \bibinfo{title}{A modular optimal formation control scheme of
  multi-agent systems with application to multiple mobile robots}.
\newblock \bibinfo{journal}{IEEE Transactions on Industrial Electronics}
  \bibinfo{volume}{69}, \bibinfo{pages}{9331--9341}.
\bibitem[{Wang et~al.(2020c)Wang, Wang and Li}]{modular_wang2020}
\bibinfo{author}{Wang, X.}, \bibinfo{author}{Wang, G.}, \bibinfo{author}{Li,
  S.}, \bibinfo{year}{2020}c.
\newblock \bibinfo{title}{Distributed finite-time optimization for integrator
  chain multiagent systems with disturbances}.
\newblock \bibinfo{journal}{IEEE Transactions on Automatic control}
  \bibinfo{volume}{65}, \bibinfo{pages}{5296--5311}.
\bibitem[{Xu et~al.(2022)Xu, He, Ho and Kurths}]{DY3-2022-1}
\bibinfo{author}{Xu, W.}, \bibinfo{author}{He, W.}, \bibinfo{author}{Ho, D.W.},
  \bibinfo{author}{Kurths, J.}, \bibinfo{year}{2022}.
\newblock \bibinfo{title}{Fully distributed observer-based consensus protocol:
  Adaptive dynamic event-triggered schemes}.
\newblock \bibinfo{journal}{Automatica} \bibinfo{volume}{139},
  \bibinfo{pages}{no.110188}.
\bibitem[{Xu et~al.(2021)Xu, Fang, Shi, Pan and Ahn}]{DY4_2019event-United2}
\bibinfo{author}{Xu, Y.}, \bibinfo{author}{Fang, M.}, \bibinfo{author}{Shi,
  P.}, \bibinfo{author}{Pan, Y.}, \bibinfo{author}{Ahn, C.K.},
  \bibinfo{year}{2021}.
\newblock \bibinfo{title}{Multileader multiagent systems containment control
  with event-triggering}.
\newblock \bibinfo{journal}{IEEE Transactions on Systems, Man, and Cybernetics:
  Systems} \bibinfo{volume}{51}, \bibinfo{pages}{1642--1651}.
\bibitem[{Xu et~al.(2019)Xu, Wu, Pan, Ahn and Yan}]{6bu-3}
\bibinfo{author}{Xu, Y.}, \bibinfo{author}{Wu, Z.}, \bibinfo{author}{Pan, Y.},
  \bibinfo{author}{Ahn, C.K.}, \bibinfo{author}{Yan, H.}, \bibinfo{year}{2019}.
\newblock \bibinfo{title}{Consensus of linear multiagent systems with
  input-based triggering condition}.
\newblock \bibinfo{journal}{IEEE Transactions on Systems, Man, and Cybernetics:
  Systems} \bibinfo{volume}{49}, \bibinfo{pages}{2308--2317}.
\bibitem[{Yoo(2013)}]{nonlinear_Sun2013}
\bibinfo{author}{Yoo, S.J.}, \bibinfo{year}{2013}.
\newblock \bibinfo{title}{Distributed adaptive containment control of uncertain
  nonlinear multi-agent systems in strict-feedback form}.
\newblock \bibinfo{journal}{Automatica} \bibinfo{volume}{49},
  \bibinfo{pages}{2145--2153}.
\bibitem[{Zhang et~al.(2020)Zhang, Wang, Peng, Li and Liu}]{DY2_2020event-T1}
\bibinfo{author}{Zhang, Y.}, \bibinfo{author}{Wang, D.}, \bibinfo{author}{Peng,
  Z.}, \bibinfo{author}{Li, T.}, \bibinfo{author}{Liu, L.},
  \bibinfo{year}{2020}.
\newblock \bibinfo{title}{Event-triggered iss-modular neural network control
  for containment maneuvering of nonlinear strict-feedback multi-agent
  systems}.
\newblock \bibinfo{journal}{Neurocomputing} \bibinfo{volume}{377},
  \bibinfo{pages}{314--324}.
\bibitem[{Zhu et~al.(2022)Zhu, Liu, Gu, Luo and L{\"u}}]{DY6-2021formation-2}
\bibinfo{author}{Zhu, G.}, \bibinfo{author}{Liu, K.}, \bibinfo{author}{Gu, H.},
  \bibinfo{author}{Luo, W.}, \bibinfo{author}{L{\"u}, J.},
  \bibinfo{year}{2022}.
\newblock \bibinfo{title}{Observer-based event-triggered formation control of
  multi-agent systems with switching directed topologies}.
\newblock \bibinfo{journal}{IEEE Transactions on Circuits and Systems I:
  Regular Papers} \bibinfo{volume}{69}, \bibinfo{pages}{1323--1332}.
\bibitem[{Zou and Xiang(2017)}]{DY0-2017-4}
\bibinfo{author}{Zou, W.}, \bibinfo{author}{Xiang, Z.}, \bibinfo{year}{2017}.
\newblock \bibinfo{title}{Event-triggered distributed containment control of
  heterogeneous linear multi-agent systems by an output regulation approach}.
\newblock \bibinfo{journal}{International Journal of Systems Science}
  \bibinfo{volume}{48}, \bibinfo{pages}{2041--2054}.
\bibitem[{Zou and Xiang(2019)}]{DY4_2019event-United1}
\bibinfo{author}{Zou, W.}, \bibinfo{author}{Xiang, Z.}, \bibinfo{year}{2019}.
\newblock \bibinfo{title}{Event-triggered containment control of second-order
  nonlinear multi-agent systems}.
\newblock \bibinfo{journal}{Journal of the Franklin Institute}
  \bibinfo{volume}{356}, \bibinfo{pages}{10421--10438}.
\bibitem[{Zuo et~al.(2017)Zuo, Song, Lewis and Davoudi}]{2017output}
\bibinfo{author}{Zuo, S.}, \bibinfo{author}{Song, Y.}, \bibinfo{author}{Lewis,
  F.L.}, \bibinfo{author}{Davoudi, A.}, \bibinfo{year}{2017}.
\newblock \bibinfo{title}{Output containment control of linear heterogeneous
  multi-agent systems using internal model principle}.
\newblock \bibinfo{journal}{IEEE Transactions on Cybernetics}
  \bibinfo{volume}{47}, \bibinfo{pages}{2099--2109}.
\bibitem[{Zuo et~al.(2018)Zuo, Song, Lewis and Davoudi}]{DY6_2018containment}
\bibinfo{author}{Zuo, S.}, \bibinfo{author}{Song, Y.}, \bibinfo{author}{Lewis,
  F.L.}, \bibinfo{author}{Davoudi, A.}, \bibinfo{year}{2018}.
\newblock \bibinfo{title}{Adaptive output containment control of heterogeneous
  multi-agent systems with unknown leaders}.
\newblock \bibinfo{journal}{Automatica} \bibinfo{volume}{92},
  \bibinfo{pages}{235--239}.

\end{thebibliography}


\begin{thebibliography}{xx}

\harvarditem[Aranovskiy {\em et al.}]{Aranovskiy {\em et al.}}{2016}{7526771}
Aranovskiy, S., Alexey Bobtsov, Romeo Ortega and Anton Pyrkin  (2016).
  Parameters estimation via dynamic regressor extension and mixing. In `2016
  American Control Conference (ACC)'. pp.~6971--6976.

\harvarditem[Berberich {\em et al.}]{Berberich {\em et al.}}{2021}{9109670}
Berberich, J., Johannes Köhler, Matthias~A. Müller and Frank Allgöwer
  (2021). `Data-driven model predictive control with stability and robustness
  guarantees'. {\em IEEE Transactions on Automatic Control} {\bf
  66}(4),~1702--1717.

\harvarditem[Bombois {\em et al.}]{Bombois {\em et al.}}{2006}{BOMBOIS20061651}
Bombois, X., G.~Scorletti, M.~Gevers, P.M.J. {Van den Hof} and R.~Hildebrand
  (2006). `Least costly identification experiment for control'. {\em
  Automatica} {\bf 42}(10),~1651--1662.

\harvarditem[Bongard {\em et al.}]{Bongard {\em et al.}}{2023}{9744574}
Bongard, J., Julian Berberich, Johannes Köhler and Frank Allgöwer  (2023).
  `Robust stability analysis of a simple data-driven model predictive control
  approach'. {\em IEEE Transactions on Automatic Control} {\bf
  68}(5),~2625--2637.

\harvarditem[Chen {\em et al.}]{Chen {\em et al.}}{2014}{W6578135}
Chen, W., Changyun Wen, Shaoyong Hua and Changyin Sun  (2014). `Distributed
  cooperative adaptive identification and control for a group of
  continuous-time systems with a cooperative {PE} condition via consensus'.
  {\em IEEE Transactions on Automatic Control} {\bf 59}(1),~91--106.

\harvarditem[De~Persis and Tesi]{De~Persis and Tesi}{2020}{8933093}
De~Persis, C. and Pietro Tesi  (2020). `Formulas for data-driven control:
  Stabilization, optimality, and robustness'. {\em IEEE Transactions on
  Automatic Control} {\bf 65}(3),~909--924.

\harvarditem[{De Persis} and Tesi]{{De Persis} and Tesi}{2021}{DEPERSIS2021285}
{De Persis}, C. and Pietro Tesi  (2021). `Designing experiments for data-driven
  control of nonlinear systems'. {\em IFAC-PapersOnLine} {\bf 54}(9),~285--290.

\harvarditem[Dhar {\em et al.}]{Dhar {\em et al.}}{2022}{DHAR2022100672}
Dhar, A., Sayan~Basu Roy and Shubhendu Bhasin  (2022). `Initial excitation
  based discrete-time multi-model adaptive online identification'. {\em
  European Journal of Control} {\bf 68},~100672.
\newblock 2022 European Control Conference Special Issue.

\harvarditem[Duan and Yu]{Duan and Yu}{2013}{duan2013lmis}
Duan, G.-R. and Hai-Hua Yu  (2013). {\em LMIs in Control Systems: Analysis,
  Design and Applications}. CRC press.

\harvarditem[Gevers {\em et al.}]{Gevers {\em et al.}}{2009}{5325719}
Gevers, M., Alexandre~Sanfelice Bazanella, Xavier Bombois and Ljubisa Miskovic
  (2009). `Identification and the information matrix: How to get just
  sufficiently rich?'. {\em IEEE Transactions on Automatic Control} {\bf
  54}(12),~2828--2840.

\harvarditem[Hamaya {\em et al.}]{Hamaya {\em et
  al.}}{2021}{doi:10.1177/0278364919853618}
Hamaya, M., Takamitsu Matsubara, Tatsuya Teramae, Tomoyuki Noda and Jun
  Morimoto  (2021). `Design of physical user–robot interactions for model
  identification of soft actuators on exoskeleton robots'. {\em The
  International Journal of Robotics Research} {\bf 40}(1),~397--410.

\harvarditem[Hjalmarsson]{Hjalmarsson}{2005}{HJALMARSSON2005393}
Hjalmarsson, H.  (2005). `From experiment design to closed-loop control'. {\em
  Automatica} {\bf 41}(3),~393--438.

\harvarditem[Ioannou and Fidan]{Ioannou and Fidan}{2006}{ioannou2006adaptive}
Ioannou, P. and Baris Fidan  (2006). {\em Adaptive Control Tutorial}. SIAM.

\harvarditem[Kang and You]{Kang and You}{2023}{KANG2023111130}
Kang, S. and Keyou You  (2023). `Minimum input design for direct data-driven
  property identification of unknown linear systems'. {\em Automatica} {\bf
  156},~111130.

\harvarditem[Katayama {\em et al.}]{Katayama {\em et
  al.}}{2005}{katayama2005subspace}
Katayama, T. et~al.  (2005). {\em Subspace Methods for System Identification}.
  Vol.~1. Springer.

\harvarditem[Liu {\em et al.}]{Liu {\em et al.}}{2023}{9903320}
Liu, W., Jian Sun, Gang Wang, Francesco Bullo and Jie Chen  (2023).
  `Data-driven resilient predictive control under denial-of-service'. {\em IEEE
  Transactions on Automatic Control} {\bf 68}(8),~4722--4737.

\harvarditem[Lovera {\em et al.}]{Lovera {\em et al.}}{2000}{LOVERA20001639}
Lovera, M., Tony Gustafsson and Michel Verhaegen  (2000). `Recursive subspace
  identification of linear and non-linear {W}iener state-space models'. {\em
  Automatica} {\bf 36}(11),~1639--1650.

\harvarditem[Markovsky]{Markovsky}{2012}{markovsky2012low}
Markovsky, I.  (2012). {\em Low Rank Approximation: Algorithms, Implementation,
  Applications}. Vol. 906. Springer.

\harvarditem[Markovsky and Dörfler]{Markovsky and
  Dörfler}{2021}{MARKOVSKY202142}
Markovsky, I. and Florian Dörfler  (2021). `Behavioral systems theory in
  data-driven analysis, signal processing, and control'. {\em Annual Reviews in
  Control} {\bf 52},~42--64.

\harvarditem[Markovsky and Dörfler]{Markovsky and Dörfler}{2023}{9904308}
Markovsky, I. and Florian Dörfler  (2023). `Identifiability in the behavioral
  setting'. {\em IEEE Transactions on Automatic Control} {\bf
  68}(3),~1667--1677.

\harvarditem[Markovsky and Pintelon]{Markovsky and Pintelon}{2015}{7098430}
Markovsky, I. and Rik Pintelon  (2015). `Identification of linear
  time-invariant systems from multiple experiments'. {\em IEEE Transactions on
  Signal Processing} {\bf 63}(13),~3549--3554.

\harvarditem[Markovsky {\em et al.}]{Markovsky {\em et
  al.}}{2005}{MARKOVSKY2005755}
Markovsky, I., Jan~C. Willems, Paolo Rapisarda and Bart~L.M. {De Moor}  (2005).
  `Algorithms for deterministic balanced subspace identification'. {\em
  Automatica} {\bf 41}(5),~755--766.

\harvarditem[Nortmann and Mylvaganam]{Nortmann and Mylvaganam}{2023}{10124991}
Nortmann, B. and Thulasi Mylvaganam  (2023). `Direct data-driven control of
  linear time-varying systems'. {\em IEEE Transactions on Automatic Control}
  {\bf 68}(8),~4888--4895.

\harvarditem[Ortega {\em et al.}]{Ortega {\em et al.}}{2021}{9121700}
Ortega, R., Stanislav Aranovskiy, Anton~A. Pyrkin, Alessandro Astolfi and
  Alexey~A. Bobtsov  (2021). `New results on parameter estimation via dynamic
  regressor extension and mixing: Continuous and discrete-time cases'. {\em
  IEEE Transactions on Automatic Control} {\bf 66}(5),~2265--2272.

\harvarditem[Park and Ikeda]{Park and Ikeda}{2009}{PARK20091265}
Park, U.~S. and Masao Ikeda  (2009). `Stability analysis and control design of
  {LTI} discrete-time systems by the direct use of time series data'. {\em
  Automatica} {\bf 45}(5),~1265--1271.

\harvarditem[Tan and Hoo]{Tan and Hoo}{2015}{7409511}
Tan, R.~H. and Landon Y.~H. Hoo  (2015). Dc-dc converter modeling and
  simulation using state space approach. In `2015 IEEE Conference on Energy
  Conversion (CENCON)'. pp.~42--47.

\harvarditem[van Waarde]{van Waarde}{2022}{9406124}
van Waarde, H.~J.  (2022). `Beyond persistent excitation: Online experiment
  design for data-driven modeling and control'. {\em IEEE Control Systems
  Letters} {\bf 6},~319--324.

\harvarditem[van Waarde {\em et al.}]{van Waarde {\em et al.}}{2020{\em
  a}}{9062331}
van Waarde, H.~J., Claudio De~Persis, M.~Kanat Camlibel and Pietro Tesi
  (2020{\em a}). `Willems’ fundamental lemma for state-space systems and its
  extension to multiple datasets'. {\em IEEE Control Systems Letters} {\bf
  4}(3),~602--607.

\harvarditem[van Waarde {\em et al.}]{van Waarde {\em et al.}}{2020{\em
  b}}{8960476}
van Waarde, H.~J., Jaap Eising, Harry~L. Trentelman and M.~Kanat Camlibel
  (2020{\em b}). `Data informativity: A new perspective on data-driven analysis
  and control'. {\em IEEE Transactions on Automatic Control} {\bf
  65}(11),~4753--4768.

\harvarditem[Van~Waarde {\em et al.}]{Van~Waarde {\em et al.}}{2023}{10317631}
Van~Waarde, H.~J., Jaap Eising, M.~Kanat Camlibel and Harry~L. Trentelman
  (2023). `The informativity approach: To data-driven analysis and control'.
  {\em IEEE Control Systems Magazine} {\bf 43}(6),~32--66.

\harvarditem[Venkatasubramanian {\em et al.}]{Venkatasubramanian {\em et
  al.}}{2025}{10601334}
Venkatasubramanian, J., Johannes Köhler, Julian Berberich and Frank Allgöwer
  (2025). `Sequential learning and control: Targeted exploration for robust
  performance'. {\em IEEE Transactions on Automatic Control} {\bf
  70}(1),~307--322.

\harvarditem[Willems {\em et al.}]{Willems {\em et al.}}{2005}{WILLEMS2005325}
Willems, J.~C., Paolo Rapisarda, Ivan Markovsky and Bart~L.M. {De Moor}
  (2005). `A note on persistency of excitation'. {\em Systems \& Control
  Letters} {\bf 54}(4),~325--329.

\harvarditem[Xie and Guo]{Xie and Guo}{2018}{doi:10.1137/16M1106791}
Xie, S. and Lei Guo  (2018). `Analysis of normalized least mean squares-based
  consensus adaptive filters under a general information condition'. {\em SIAM
  Journal on Control and Optimization} {\bf 56}(5),~3404--3431.

\harvarditem[Yang {\em et al.}]{Yang {\em et al.}}{2024}{YANG2024100986}
Yang, Y.-N., Jie Jin, Li-Tao Zhu, Yin-Ning Zhou and Zheng-Hong Luo  (2024).
  `Runaway criteria for predicting the thermal behavior of chemical reactors'.
  {\em Current Opinion in Chemical Engineering} {\bf 43},~100986.

\harvarditem[Yu {\em et al.}]{Yu {\em et al.}}{2021}{9682952}
Yu, Y., Shahriar Talebi, Henk~J. van Waarde, Ufuk Topcu, Mehran Mesbahi and
  Behcet Acıkmese  (2021). On controllability and persistency of excitation in
  data-driven control: Extensions of {W}illems’ fundamental lemma. In `2021
  60th IEEE Conference on Decision and Control (CDC)'. pp.~6485--6490.

\end{thebibliography}
\bibliographystyle{automatica1} 

\end{document}